\documentclass[prd, showpacs, floatfix,onecolumn,nofootinbib]{revtex4}
\usepackage{amsmath}
\usepackage{latexsym}
\usepackage{graphicx,epsfig,color}
\usepackage{amssymb}
\usepackage{subfigure}
\usepackage{epstopdf}
\usepackage[colorlinks=true,citecolor=blue,linkcolor=blue,urlcolor=black]{hyperref}
\begin{document}

\title{\textbf{Extended phase space thermodynamics for Lovelock black holes with non-maximally symmetric horizons}}
\author{N. Farhangkhah$^1$\footnote{email address:
farhangkhah@iaushiraz.ac.ir} and Z. Dayyani$^2$}
\affiliation{$^1$ Department of Physics, Shiraz Branch, Islamic Azad University, Shiraz 71993,
Iran\\
       $^2$ Physics Department and Biruni Observatory, College of Sciences, Shiraz
University, Shiraz 71454, Iran}

\begin{abstract}
We study thermodynamics and critical behaviors of higher-dimensional Lovelock black holes
with non-maximally symmetric horizons in the canonical ensemble of extended phase space.
The effects from non-constancy of the horizon of the black hole via appearing
two chargelike parameters in thermodynamic quantities of third-order Lovelock black hole are investigated.
We find that Ricci flat black holes with nonconstant curvature horizon show critical behavior. This is an
interesting feature that is not seen for any kind of black hole in Einstein or
Lovelock gravity in the literature. We examine how various interesting thermodynamic phenomena such as
standard first-order small-large black hole phase transition, a reentrant phase transition, or zeroth order
phase transition happen for Ricci flat, spherical, or hyperbolic black holes with nonconstant curvature horizon
depending on the values
of Lovelock coefficient and chargelike parameters.
While for a spherical black hole of third order Lovelock gravity with constant curvature horizon phase transition is observed
only for $7\leq d \leq11$, for our solution criticality and phase transition exist in every dimension.
With a proper choice of the free parameters, a large-small-large black hole phase transition occurs.
This process is accompanied
by a finite jump of the Gibbs free energy referred to as a zeroth-order phase transition. For the case $\kappa=-1$
a novel behavior is found for which three critical points could exist.

\end{abstract}

\maketitle

\section{Introduction}

\bigskip Einstein's theory of general relativity, is the most successful
theory of gravity. At very high energies close to the Planck scale, higher
order curvature terms can no longer be neglected. Over the past years,
motivated by Superstring/M-theory \cite{string}, higher dimensional gravity
has been a prevailing subject of study. The most famous theory that
generalizes general relativity in higher dimensions is Lovelock theory \cite%
{Lovelock}. This theory keeps the order of the field equations down to
second order in derivatives. The Lovelock Lagrangian consists of a sum of
dimensionally extended Euler densities, and the second-order field equations
in this model give rise to the ghost-free nature of the theory. The best
testing ground for any modified theory of gravity will be to search for
black hole solutions. Investigating different aspects of black hole physics has
raised many interests.

Since it was found that black holes in classical general theory of
relativity obey laws that are analogous to the laws of thermodynamics, a lot
of attentions have been attracted to study the thermodynamic properties of
the black holes. First, Hawking proposed that the area of the black hole
event horizon never decreases. This is analogous to the second law of
thermodynamics with area of event horizon playing the role of entropy in
thermodynamics. After that the black hole entropy was introduced as
proportional to the black hole surface area in Planck units by Bekenstein
\cite{Beken}. Hawking also suggested the temperature for the black hole \cite%
{Hawking}, and proposed that like any other hot body, the black hole
radiates in analogy to the zeroth law of thermodynamics. Thus the black hole
can be considered to be a thermodynamic object and study of black hole
thermodynamics has provided interesting information about the underlying
structure of the spacetime. The discovery of the famous Hawking-Page phase
transition in Schwarzschild Anti de Sitter (AdS) black holes \cite{Haw-Page}
was a begin to a wide area of researches in the context of black hole
thermodynamics.

\bigskip\ In thermal systems, Van der Waals equation modifies the equation
of state for an ideal gas to one that approximates the behavior of real
fluids. Through studying the thermodynamics of charged black holes, it was
found that the first order small-large phase transition for charged black
hole in AdS space is quite similar to the liquid-gas change of phase
occurring in Van der Waals fluids \cite{Chamb1, Chamb2}. It was showed that $%
Q-\Phi$ diagram of the charged black holes is similar to the $P-V$
diagram of the van der Waals system. Researches in this regards, led to the
assumption of an \textit{extended thermodynamic phase space}. In this
framework, black hole thermodynamics is studied in the asymptotic AdS space
with a negative cosmological constant $\Lambda$. The cosmological constant
is represented as a pressure $(\Lambda=-P/8\pi),$ and the
thermodynamically conjugate variable is the thermodynamic volume \cite{Dolan1, Kastor1, Cvetic}.
There exist good reasons why the variation of
$\Lambda$ should be included in thermodynamic considerations \cite {Gibb, Crei}:
Firstly, in theories where
physical constants such as Yukawa couplings, gauge coupling
constants, or the cosmological
constant are not fixed a priori, but arise as vacuum expectation
values and hence can vary, it is
natural to include variations of these ‘constants’ in the
thermodynamic formulae such as the first law. In fact such ‘constants’ are typically to be thought of
as the values at infinity of scalar fields.
Secondly, in the presence
of a cosmological constant the first law of black hole
thermodynamics becomes inconsistent with the Smarr relation \cite{Smarr} unless
the variation of $\Lambda$ is included in the first law and then the black hole
mass $M$ is identified with enthalpy rather than internal
energy. One can also define other thermodynamic quantities of black holes such as
adiabatic compressibility, specific heat at constant pressure, or even the
speed of sound \cite{Dolan2, Dolan3}.
In \cite{Karch} it is shown that from simple field theoretic considerations a universal Smarr formula
emerges in holographic descriptions of black holes with large $N$ duals and considering $\Lambda$ as a dynamical variable can
be understood from the point of view of the dual holographic field theory.

From another point of view, one can introduce a new gauge field in
the Lagrangian in which $\Lambda$ appears as the conserved charge associated with the global part
of the gauge symmetry of this gauge field. This formulation
brings a new perspective to $\Lambda$ as a parameter of the solution, and can naturally contribute to the first law of black hole
thermodynamics just like other solution parameters like mass, entropy, angular momentum, electric charge etc. \cite{Hajian}

The analysis of the $P-V$ critical behaviors in the extended phase space has
been under study extensively and generalized to higher dimensional charged
black holes \cite{Kubiz, Belhaj1, Spall}, rotating black holes \cite{Chabab, Alta1},
and black holes with Born-Infeld filed \cite{Guna}.
If a monotonic variation of any thermodynamic quantity results in two (or more)
phase transitions such that the final state is macroscopically similar to the initial state,
the system undergoes the reentrant phase transition \cite{Nara,Alta1}.
The reentrant phase transition was found for the
four-dimensional Born-Infeld-AdS black hole spacetimes, \cite{Guna},
and for the black holes of third-order Lovelock gravity \cite{Frassino}.
The situation is accompanied by a discontinuity in the global minimum
of the Gibbs free energy, referred to as a zeroth-order
phase transition and seen in superfluidity and superconductivity \cite{Maslov}.

Corrections to black hole thermodynamics from higher-curvature terms in
Lovelock theory have revealed interesting features. In Lovelock theory the
entropy is given by a complicated relationship depending on higher-curvature
terms, and is no longer proportional to the area of the horizon \cite{Iyer}. 
Both the first law and the associated Smarr formula in an
extended phase space were obtained exploiting the Killing potential formalism \cite{Kastor2}.
It is also proposed to introduce a quantity conjugated to the Lovelock coefficient in
the first law of black hole thermodynamics and in the Smarr relation. 
$P-V$ criticality has been searched for Gauss-Bonnet \cite{Cai, Zou, Sheykhi, Hendi} and
third-order Lovelock \cite{Xu, Mo, Belhaj2} black holes. It was shown in
\cite{Cai} that $P-V$ criticality can be observed for spherical Gauss-Bonnet
black holes even when charge is absent. For third order Lovelock gravity, it is
found that for $\kappa=1$ only for dimensions $7\leq d\leq 11$ critical points
exists. In all the cases mentioned above, for $\kappa=0$ there is no critical
point in the extended thermodynamic phase space. All the works mentioned above
have considered black hole with maximally symmetric horizons. In Ref.
\cite{Dotti} a novel class of black hole solution is derived with
nonconstant curvature horizon in Lovelock gravity. The properties of
this kind of black hole are investigated in second and third order
Lovelock gravity \cite{Maeda, Farhang, Farhang2, Ohashi, Ray2}.
A noteworthy change
when considering nonconstant curvature horizon is that new chargelike
parameters appear in the metric function with the advantage of higher
curvature terms, and modify the properties of the black holes. Specially,
Ricci flat solutions of this kind of black holes show interesting features
and this motivates us to investigate the effects of non constancy of the
horizon on the $P-V$ criticality of such black holes.

Our paper is organized as follows. We begin in Sec. \ref{thermo} by reviewing the
solutions of Lovelock gravity with non-constant curvature horizons and the extended
phase space thermodynamics in Lovelock theory is discussed. In Secs. \ref{criticalityf},
\ref{criticalitys} and \ref{criticalityh},
specifying to black holes of 3-order Lovelock gravity, critical behavior of the black hole
with nonconstant horizon but constant sectional curvatures $\kappa=0,\pm 1$ is studied,
and we will see how criticality and phase transition can occur in various
cases. Also we will calculate the critical exponents
for these black holes and show that they are in the same university class as
the van der Waals gas. Our results are summarized in the concluding section \ref{conclusion}.

\bigskip

\section{\protect\bigskip Extended thermodynamics of nonmaximally symmetric Lovelock AdS black holes%
} \label{thermo}

To start, we consider the physical action describing Lovelock gravity which
is in the following form:
\begin{equation}
I=\int_{\mathcal{M}}d^{d}x\sqrt{-g}\left( -2\Lambda +\sum_{p=1}^{\overline{p}%
}\alpha _{p}\mathcal{L}^{(p)}\right) .  \label{Act}
\end{equation}%
where $\Lambda $ is the cosmological constant and $\alpha _{p}$'s are the
Lovelock coupling constants with the choose of $\alpha _{1}=1$. The Einstein
term $\mathcal{L}^{(1)}$ equals to $R$ and the second order Lovelock term is
$\mathcal{L}^{(2)}=R_{\mu \nu \gamma \delta }R^{\mu \nu \gamma \delta
}-4R_{\mu \nu }R^{\mu \nu }+R^{2}.$ Also $\mathcal{L}^{(3)}$ is the third
order Lovelock Lagrangian which is described as
\begin{eqnarray}
\mathcal{L}^{(3)} &=&2R^{\mu \nu \sigma \kappa }R_{\sigma \kappa \rho \tau
}R_{\phantom{\rho \tau }{\mu \nu }}^{\rho \tau }+8R_{\phantom{\mu
\nu}{\sigma \rho}}^{\mu \nu }R_{\phantom {\sigma \kappa} {\nu \tau}}^{\sigma
\kappa }R_{\phantom{\rho \tau}{ \mu \kappa}}^{\rho \tau }+24R^{\mu \nu
\sigma \kappa }R_{\sigma \kappa \nu \rho }R_{\phantom{\rho}{\mu}}^{\rho }
\notag \\
&&+3RR^{\mu \nu \sigma \kappa }R_{\sigma \kappa \mu \nu }+24R^{\mu \nu
\sigma \kappa }R_{\sigma \mu }R_{\kappa \nu }+16R^{\mu \nu }R_{\nu \sigma
}R_{\phantom{\sigma}{\mu}}^{\sigma }-12RR^{\mu \nu }R_{\mu \nu }+R^{3}.
\label{ToL}
\end{eqnarray}

We start with the following metric%
\begin{equation}
ds^{2}=-f(r)dt^{2}+f^{-1}(r)dr^{2}+r^{2}\gamma _{ij}(z)dz^{i}dz^{j},
\label{metric1}
\end{equation}%
which is a warped product of a $2$-dimensional Riemannian submanifold $M^{2}$
and an $(d-2)$-dimensional submanifold $K^{(d-2)}$. In this relation $i,j$
go from $2,...,d-1$. The submanifold $K^{(d-2)}$ with the unit metric $%
\gamma _{ij}$ is assumed to be an Einstein manifold with nonconstant
curvature but having a constant Ricci scalar being

\begin{equation}
\widetilde{R}=\kappa (d-2)(d-3),\text{ \ }  \label{Ricci Sca}
\end{equation}%
with $\kappa $ being the sectional curvature. For the tensor components of
the submanifold \ $K^{(d-2)}$ a tilde is used. The Ricci and Riemann tensors
of the Einstein manifold are

\begin{eqnarray}
\text{\ \ \ \ \ \ \ }\widetilde{R}_{ij} &=&\kappa (d-3)\gamma _{ij},
\label{Ricci Ten} \\
\widetilde{{R}}{_{ij}}^{kl} &=&\widetilde{{C}}{_{ij}}^{kl}+\kappa ({\delta
_{i}}^{k}{\delta _{j}}^{l}-{\delta _{i}}^{l}{\delta _{j}}^{k})\text{\ },
\label{Riemm Ten}
\end{eqnarray}%
where $\widetilde{{C}}_{ij}^{kl}$ is the Weyl tensor of $K^{(d-2)}$.

Choosing $\overline{p}=3$ in the field equation, for the metric (\ref%
{metric1}) to be a solution of field equations in third order Lovelock
theory in vacuum, it would suffice that the Weyl tensor of the horizon
satisfies the following constraints
\begin{equation}
\sum_{kln}\widetilde{{C}}{_{ki}}^{nl}\widetilde{{C}}{_{nl}}^{kj}=\frac{1}{d}{%
\delta _{i}}^{j}\sum_{kmpq}\widetilde{{C}}{_{km}}^{pq}\widetilde{{C}}{_{pq}}%
^{km}\equiv \eta _{2}{\delta _{i}}^{j},  \label{eta2}
\end{equation}

\begin{eqnarray}
&&\sum_{klnmp}2(4\widetilde{{C}}{^{nm}}_{pk}\widetilde{{C}}{^{kl}}_{ni}%
\widetilde{{C}}{^{pj}}_{ml}-\widetilde{{C}}{^{pm}}_{ni}\widetilde{C}^{jnkl}%
\widetilde{C}_{klpm})  \notag \\
&=&\frac{2}{d}{\delta _{i}}^{j}\sum_{klmpqr}\left( 4\widetilde{{C}}{^{qm}}%
_{pk}\widetilde{{C}}{^{kl}}_{qr}\widetilde{{C}}{^{pr}}_{ml}-\widetilde{{C}}{%
^{pm}}_{qr}\widetilde{C}^{rqkl}\widetilde{C}_{klpm}\right)  \notag \\
&\equiv &\eta _{3}{\delta _{i}}^{j}.  \label{eta3}
\end{eqnarray}%
The first constraint was originally introduced by Dotti and Gleiser in \cite{Dotti}
and is due to the Gauss-Bonnet term, and the second one which is
dictated by the third order Lovelock term, is obtained in \cite{Farhang}.
These two new chargelike parameters appear in the metric function with the advantage of higher
curvature terms, and modify the properties of the black holes.

\bigskip Considering the case%
\begin{eqnarray}
\alpha _{2} &=&\frac{\alpha }{(d-3)(d-4)} \\
\alpha _{3} &=&\frac{\alpha ^{2}}{72\binom{n-2}{4}},
\end{eqnarray}%
the metric function $f(r)$ is given by \cite{Farhang}
\begin{eqnarray}
f(r) &=&\kappa +\frac{r^{2}}{\alpha }\left\{ 1+\left( j(r)\pm \sqrt{\gamma
(r)+j^{2}(r)}\right) ^{1/3}-\gamma (r)^{1/3}\left( j(r)\pm \sqrt{\gamma
(r)+j^{2}(r)}\right) ^{-1/3}\right\} ,  \notag \\
j(r) &=&-\frac{1}{2}+\frac{3\alpha }{2}\left( -\frac{2\Lambda }{(d-1)(d-2)}-%
\frac{m}{r^{d-1}}+\frac{\alpha ^{2}\hat{\eta}_{3}}{3r^{6}}\right) ,\text{ \ }
\notag \\
\text{\ \ \ }\gamma (r) &=&\left( \frac{\alpha ^{2}\hat{\eta}_{2}}{r^{4}}%
\right) ^{3},  \label{fstat}
\end{eqnarray}%
where we define $\hat{\eta }_{2}=\frac{(d-6)!\eta _{2}}{(d-2)!}$\ and $
\hat{\eta }_{3}=\frac{(d-8)!\eta _{3}}{(d-2)!}$ for simplicity. Note
that $\alpha $ and $\hat{\eta }_{2}$ are positive parameters, while $%
\hat{\eta }_{3}$ can be positive or negative relating to the metric of
the spacetime. We should also mention that in order to have the effects of
non-constancy of the curvature of the horizon in third order Lovelock
gravity, $d$ should be larger than seven, since the constants $\hat{\eta
}_{2}$ and $\hat{\eta }_{3}$ are evaluating on the $(d-2)$-dimensional
boundary

In what follows we treat the (negative) cosmological constant $\Lambda$ as
thermodynamic pressure and its conjugate quantity as thermodynamic volume
\cite{Kubiz}%
\begin{equation}
P=-\frac{\Lambda }{8\pi },
\end{equation}%
\begin{equation}
V=(\frac{\partial M}{\partial P})_{S,\alpha }=\frac{\Sigma _{d-2}r_{h}^{d-1}%
}{d-1},
\end{equation}

We obtain the parameter $M$ in terms of the horizon radius $r_{h}$ by
solving $f(r)=0$ as below%
\begin{equation}
M=\frac{(d-2)\Sigma _{d-2}}{16\pi }[\frac{16\pi P}{(d-1)(d-2)}%
r_{h}^{d-1}+\kappa r_{h}^{d-3}+\alpha (\kappa ^{2}+\hat{\eta }%
_{2}]r_{h}^{d-5}+\frac{\alpha ^{2}}{3}(\kappa ^{3}+3\hat{\eta }%
_{2}\kappa +\hat{\eta }_{3})r_{h}^{d-7}]  \label{M}
\end{equation}%
which is interpreted as enthalpy rather than the internal energy of the
gravitational system. $\Sigma _{d-2}$ denotes the volume of the $(d-2)$%
-dimensional hypersurface $K^{(d-2)}.$The Hawking temperature of such black
holes, related with the surface gravity on the horizon $r=r_{h}$ is given by
\cite{Farhang}
\begin{equation}
T=\frac{\frac{16\pi P}{(d-2)}r_{h}^{6}+(d-3)\kappa r_{h}^{4}+(d-5)(\hat{%
\eta }_{2}+\kappa ^{2})\alpha r_{h}^{2}+(d-7)\frac{\alpha ^{2}}{3}(\hat{%
\eta }_{3}+3\kappa \hat{\eta }_{2}+\kappa ^{3})}{4\pi
r_{h}[r_{h}^{4}+2\kappa \alpha r_{h}^{2}+\alpha ^{2}(\hat{\eta }%
_{2}+\kappa ^{2})]},  \label{Temp}
\end{equation}%
and entropy can be derived by making use of the Wald prescription as

\begin{equation}
S=\frac{(d-2)\Sigma _{d-2}r_{h}^{d-2}}{4}[\frac{1}{(d-2)}+\frac{2\kappa
\alpha }{r_{h}^{2}(d-4)}+\frac{\alpha ^{2}(\hat{\eta }_{2}+\kappa ^{2})}{%
r_{h}^{4}(d-6)}].  \label{Entro}
\end{equation}

The first law, in the extended phase space, yields\qquad \qquad \qquad
\qquad \qquad \qquad \qquad \qquad \qquad \qquad \qquad \qquad \qquad \qquad
\qquad \qquad \qquad
\begin{equation}
dM=TdS+VdP+\mathcal{A}d\alpha  \label{First Law}
\end{equation}%
where $\mathcal{A}$ denote the quantities conjugated to the Lovelock
coefficient and is calculated as below%
\begin{eqnarray}
\mathcal{A} &=&(\frac{\partial M}{\partial \alpha })_{S,P}=\frac{(d-2)\Sigma
_{d-2}r_{h}^{d-7}}{48\pi }\{3r^{2}(\kappa ^{2}+\widehat{\eta }_{2})+2\alpha
(\kappa ^{3}+3\kappa \widehat{\eta }_{2}+\widehat{\eta }_{3})  \notag \\
&&-\frac{(\frac{2\kappa }{(d-4)}r^{2}+\frac{2\alpha (\kappa ^{2}+\widehat{%
\eta }_{2})}{(d-6)})}{r^{4}+2\kappa \alpha r^{2}+\alpha ^{2}(\kappa ^{2}+%
\widehat{\eta }_{2})}[\frac{48\pi r^{6}P}{d-2}+3\kappa
(d-3)r^{4}+3(d-5)\alpha (\kappa ^{2}+\widehat{\eta }_{2})r^{2}+(d-7)\alpha
^{2}(\kappa ^{3}+3\kappa \widehat{\eta }_{2}+\widehat{\eta }_{3})]\}.
\label{Conjucate of alpha}
\end{eqnarray}
These thermodynamical quantities satisfy the generalized Smarr relation in
the extended phase space%
\begin{equation}
M=\frac{d-2}{d-3}TS-\frac{2}{d-3}VP+\frac{2}{d-3}\mathcal{A}\alpha
\label{Smarr Relation}
\end{equation}

One can rearrange Eq. (\ref{Temp}) to get thermodynamic equation of the
state for the black hole in the following form,%
\begin{equation}
P=\frac{T}{v}-\frac{\kappa(d-3)}{\pi (d-2)v^{2}}+\frac{32\kappa\alpha T}{(d-2)^{2}v^{3}%
}-\frac{16\alpha (d-5)(\hat{\eta} _{2}+\kappa^{2})}{\pi (d-2)^{3}v^{4}}+\frac{256\alpha
^{2}T(\hat{\eta} _{2}+\kappa^{2})}{(d-2)^{4}v^{5}}-\frac{256\alpha
^{2}(d-7)(\kappa^{3}+3\kappa\hat{\eta} _{2}+\hat{\eta} _{3})}{3\pi (d-2)^{5}v^{6}},\label{EOS}
\end{equation}%
in which we have introduced the parameter%
\begin{equation}
v=\frac{4r_{h}}{(d-2)}\label{ESV}
\end{equation}

as an effective specific volume.

If we consider the Van der Waals equation given as
\begin{equation}
P=\frac{T}{v-b}-\frac{a}{v^{2}},  \label{Van}
\end{equation}%
and make use of the series expansion%
\begin{equation}
(1-\frac{b}{v})^{-1}=\sum_{n=0}(\frac{b}{v})^{n},  \label{Van, Series}
\end{equation}
it is well seen that if we keep the higher order terms in the Taylor series
expansion, the Van der Waals equation is in correspondence with the equation
of state (\ref{EOS}) including the terms which appear from Lovelock gravity.

The critical point occurs when $P=P(v)$ has an inflection point, i.e.,
\begin{equation}
\frac{\partial P}{\partial v}=0,\text{ \ \ \ \ \ \ \ }\frac{\partial ^{2}P}{%
\partial v^{2}}=0  \label{CritEq}
\end{equation}%
and $\frac{\partial ^{2}P}{\partial v^{2}}$ changes signs around each of the
solution. One of the best ways to investigate the critical behavior and
phase transition of the system is to plot the isotherm diagrams and compare
with Van der Waals liquid-gas system. In what follows we shall investigate the $P-v
$ criticality of the black hole with nonconstant horizon but constant
sectional curvatures $\kappa=0,\pm 1.$

\section{Critical behavior of Lovelock Ricci flat black holes with $\protect\kappa =0$}
\label{criticalityf}

\begin{figure}[tbp]
	\centering
	\subfigure[ $\alpha=1$, $\hat{\eta}_{2}=1$, $\hat{\eta}_{3}=-1$ and
	$d=11$]{\includegraphics[scale=0.7]{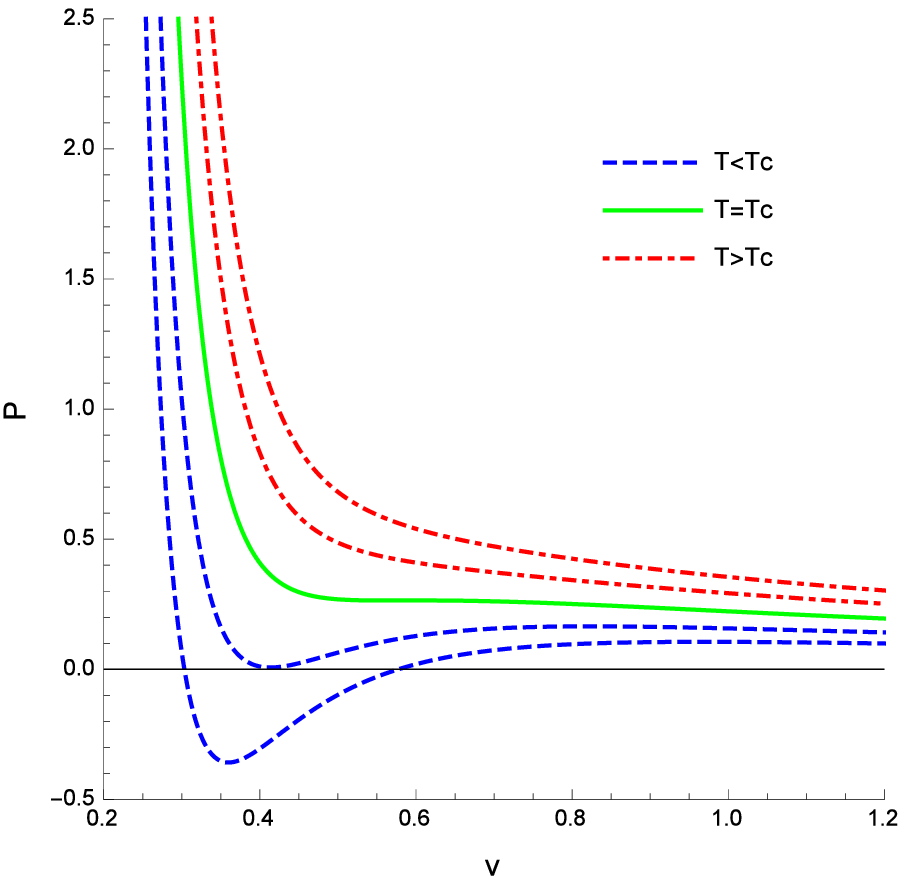}\label{Fig1c}} \hspace*{.2cm}
	\subfigure[$\alpha=1$, $\hat{\eta}_{2}=2$, $\hat{\eta}_{3}=-5$ and
	$d=8$
	]{\includegraphics[scale=0.7]{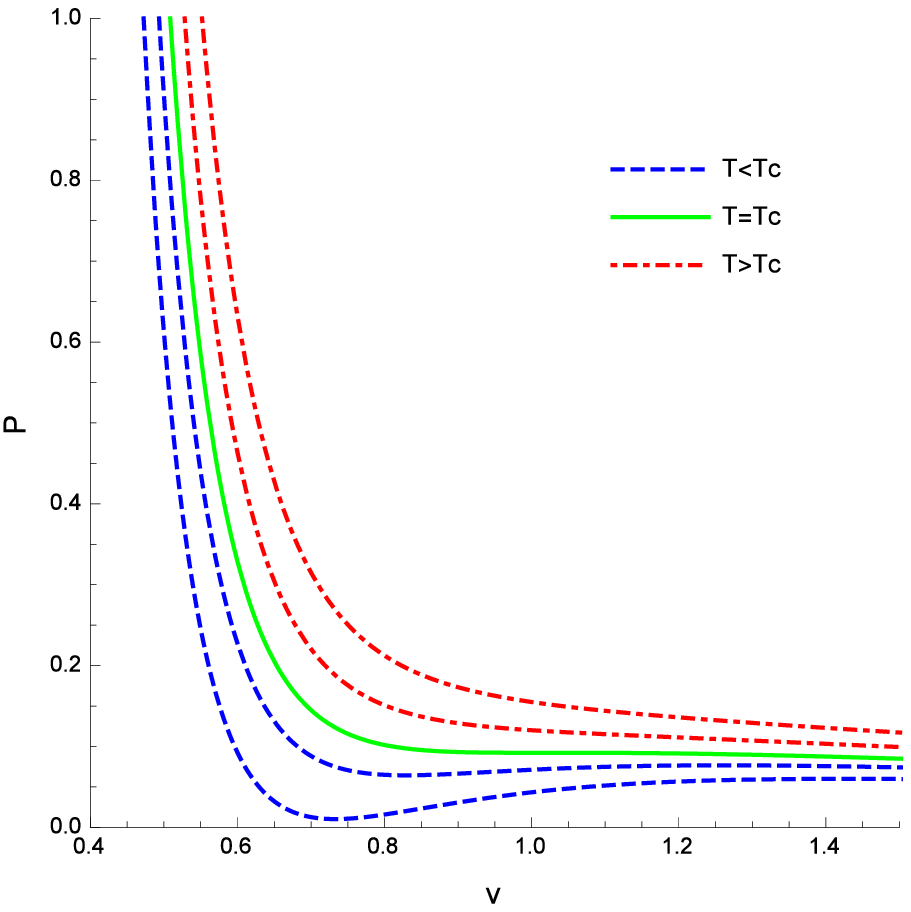}\label{Fig1b}}
	\caption{$P-v$ diagram of Lovelock black holes with $\kappa=0$ .}
	\label{Fig1}
\end{figure}

For $\kappa=0$, the equation of state can be written as
\begin{equation}
P=\frac{T}{v}-\frac{16\alpha (d-5)\hat{\eta} _{2}}{\pi (d-2)^{3}v^{4}}+\frac{%
256\alpha ^{2}\hat{\eta} _{2}T}{(d-2)^{4}v^{5}}-\frac{256\alpha ^{2}(d-7)\hat{\eta} _{3}%
}{3\pi (d-2)^{5}v^{6}},  \label{Pk0}
\end{equation}%
To obtain the critical points, if exist, we should solve Eqs. (\ref{CritEq})
which could be simplified as
\begin{equation}
x^{3}+qx+s=0\text{ \ \ \ \ \ \ \ \ \ \ \ ,\ }x=v^{2}\text{\ \ \ \ \ \ \ \ \
\ \ \ \ }  \label{EQx}
\end{equation}%
with the parameters $q$ and $s$ given by

\begin{eqnarray}
q &=&-\frac{5}{3}\hat{\eta} _{2}\mathcal{B}^{2}-3\mathcal{C}^{2},\text{ \ \ \ \ \ }%
s=\frac{35(d-7)\hat{\eta} _{3}}{27(d-5)}\mathcal{B}^{3}+2\mathcal{C}^{3}  \notag \\
\mathcal{B} &=&\frac{16\alpha }{(d-2)^{2}},\text{ \ \ \ \ \ \ \ \ \ \ \ \ \
\ \ \ \ }\mathcal{C}=\frac{40\hat{\eta} _{3}(d-7)\alpha }{9\hat{\eta}  _{2}(d-2)^{2}(d-5)}%
.  \label{a,b}
\end{eqnarray}%
As we mentioned before, $\hat{\eta} _{2}$ is a positive parameter but $\hat{\eta} _{3}$
can take an arbitrary positive or negative value. It is well known
that the multiplication of three
roots of Eq. (\ref{EQx}) is proportional to $-s$ which can be shown through a
straightforward calculation to be proportional to $-\hat{\eta} _{3}$.
Thus for negative values of $\hat{\eta} _{3},$ the multiplication of three roots of
Eq. (\ref{EQx}) is positive and thus there exists at least
one positive real root for this equation. This fact leads to the existence of at least one real root for
Eqs. (\ref{CritEq}). One should note that in the expression for $P$ which is given by the relation (%
\ref{Pk0}), the last term is dominant, as $v\rightarrow 0,$
 which is positive for negative values of $%
\hat{\eta} _{3}.$ The isotherm diagrams $P-v$ for a Ricci flat black hole are displayed
in Fig. \ref{Fig1} for two values of $d$. The plots obviously show a first order phase transition in the
system for $T<T_{c}$ which is really similar to the Van der Waals liquid-gas
system. As it is seen, for a fixed temperature lower than the critical one,
in the small radius region and large one the compression coefficient is
positive, which shows stable phases. Between them there is an unstable
phase. therefore a small/large black hole phase transition occurs.

It is worthwhile to emphasis that such a phase transition is never seen for
$\hat{\eta}_{2}=\hat{\eta} _{3}=0$. As is well known any planar black holes with constant curvature
horizon of Einstein or higher-order Lovelock gravity in an arbitrary number of spacetime dimensions
in vacuum or even in the existence of Maxwell, Born-Infeld, or dilaton fields do not admit critical behavior.
This interesting behavior is due to the existence of $\hat{\eta} _{2}$ and $\hat{\eta} _{3}$ which appear
as a result of the nonconstancy of the horizon and makes drastic changes to
the equation of state in the case $\kappa=0.$ Also there is no criticality for
$\hat{\eta}_{2}\neq0$ and $\hat{\eta}_{3}=0$. This reveals the effect of higher-curvature
terms in third-order Lovelock gravity, which cause novel changes in the
properties of the spacetime.

The solutions to Eq. (\ref{EQx}) could be written as
\begin{eqnarray}
v_{c} &=&%
\sqrt{x}=[\sqrt[3]{-\frac{s}{2}+\sqrt{(\frac{q}{3})^{3}+(\frac{s}{2})^{2}}}+%
\sqrt[3]{-\frac{s}{2}-\sqrt{(\frac{q}{3})^{3}+(\frac{s}{2})^{2}}}]^{\frac{1}{%
2}},  \label{vc} \\
T_{c} &=&\frac{(d-7)\hat{\eta} _{3}}{2(d-2)\pi \hat{\eta} _{2}v_{c}}+\frac{3}{80}\frac{%
(d-5)(d-2)}{\alpha \pi }v_{c}.  \label{Tc}
\end{eqnarray}

\begin{figure}[tbp]
\centering
\subfigure[ $\alpha=1$, $\hat{\eta}_{2}=2$, $\hat{\eta}_{3}=5$ and
	$d=8$]{\includegraphics[scale=0.7]{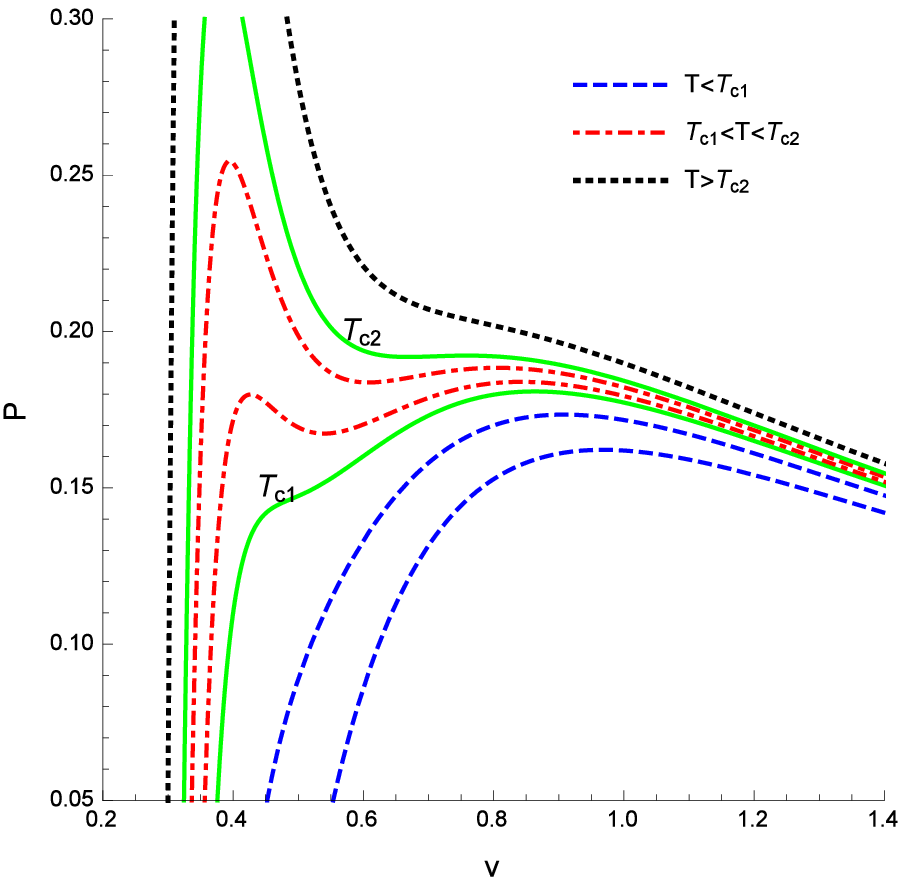}\label{Fig40a}} \hspace*{.2cm}
\subfigure[ $\alpha=1$, $\hat{\eta}_{2}=2$, $\hat{\eta}_{3}=2$ and
	$d=8$
	]{\includegraphics[scale=0.7]{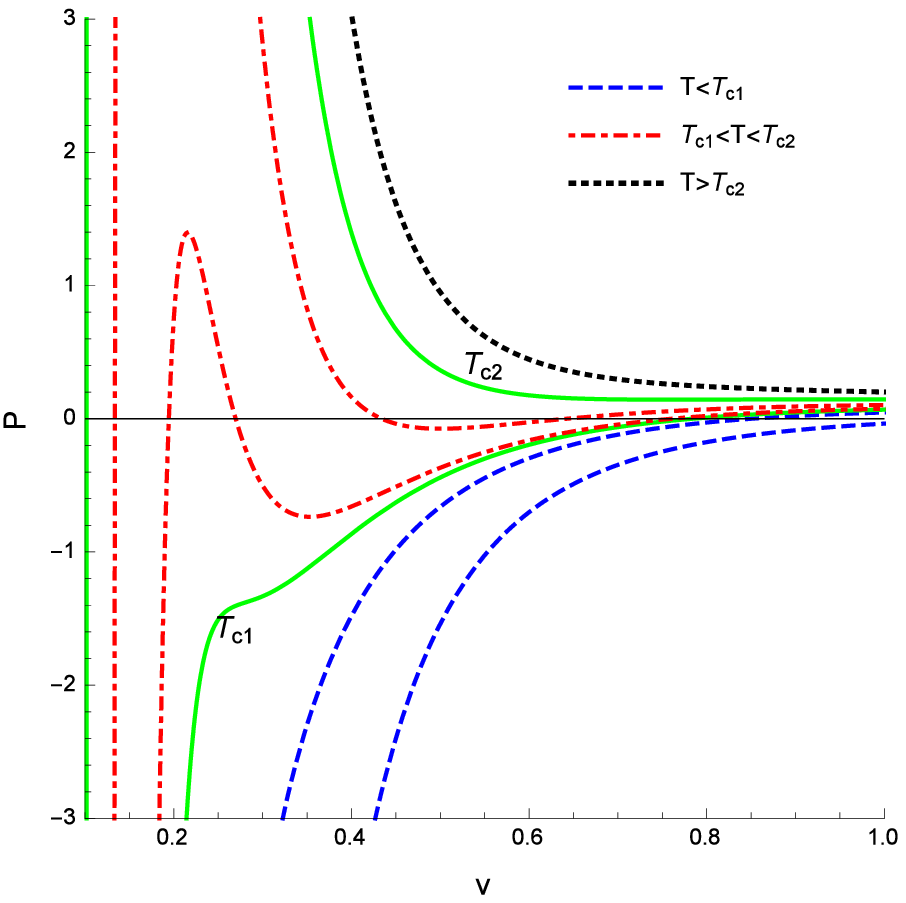}\label{Fig40b}}
\caption{$P-v$ diagram of Lovelock black holes with positive $\hat{\eta}%
_{3} $ for $\kappa=0$}
\label{Fig40}
\end{figure}

For positive values of $\hat{\eta} _{3},$ the relation (%
\ref{vc}) makes a limitation on $\hat{\eta} _{3}$ that depends on  $\alpha$, $\hat{\eta} _{2}$
and number of dimensions $d$ as
\begin{equation}
\quad \hat{\eta} _{3}<\frac{2\sqrt{3}(d-5)\hat{\eta} _{2}^{3/2}}{5(d-7)}.  \label{lim}
\end{equation}
For positive values of $\hat{\eta} _{3}$, satisfying the above constraint,
Eq. (\ref{EQx}), has at least one real root introduced as $v_{c}$. To witness the $P-v$ criticality
behavior we plot the $P-v$ diagram in Fig. \ref{Fig40}. $P-v$ diagrams in
diverse dimensions are the same and so without loss of generality we present them
for $d=8$. Fig. \ref{Fig40b} show that for some values of $\hat{\eta}_{3}$, in every dimension, there are two critical points,
one with negative (unphysical) and the other with positive pressure.
The isothermal plots in this case are quite similar with the $P-v$
diagram of Born-Infeld-AdS black holes \cite{Guna}.
As $\hat{\eta} _{3}$ is increased up to the limiting value obtained
in the relation (\ref{lim}), both values of critical pressure become positive as is
seen in \ref{Fig40a}.

\subsection{Gibbs free energy}
\begin{figure}[tbp]
	\centering
	\subfigure[$\alpha=1$, $\hat{\eta}_{2}=1$, $\hat{\eta}_{3}=-1$ and
	$d=11$]{\includegraphics[scale=0.8]{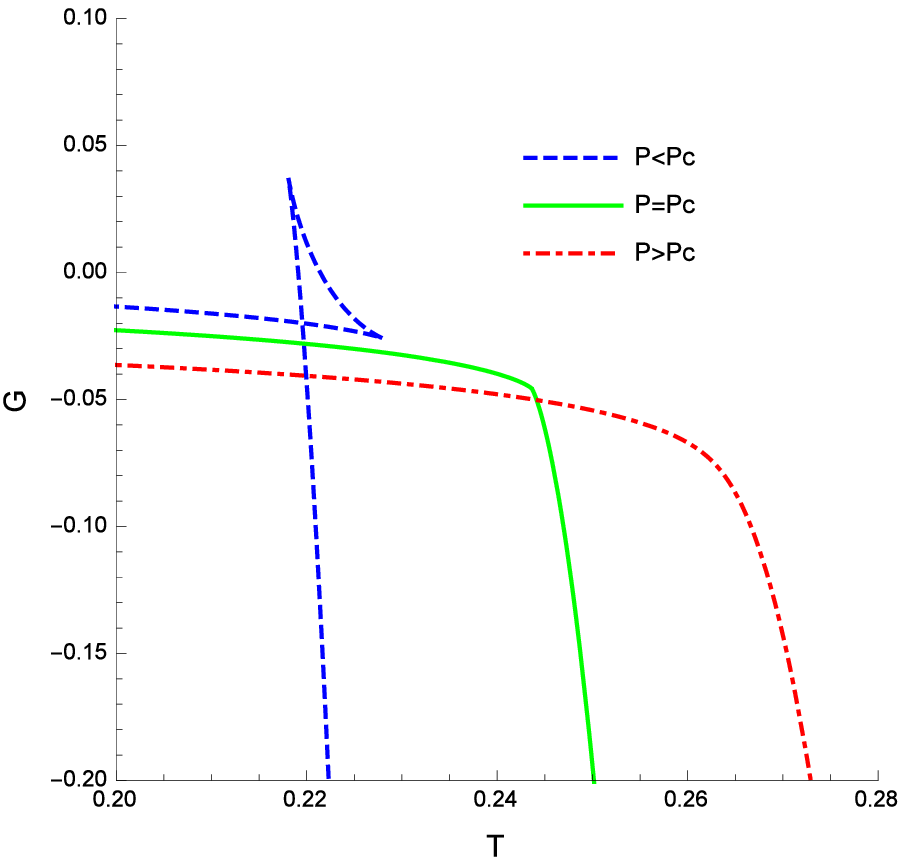}\label{Fig3a}} \hspace*{.2cm}
	\subfigure[$\alpha=1$, $\hat{\eta}_{2}=2$, $\hat{\eta}_{3}=-5$ and
	$d=8$]{\includegraphics[scale=0.8]{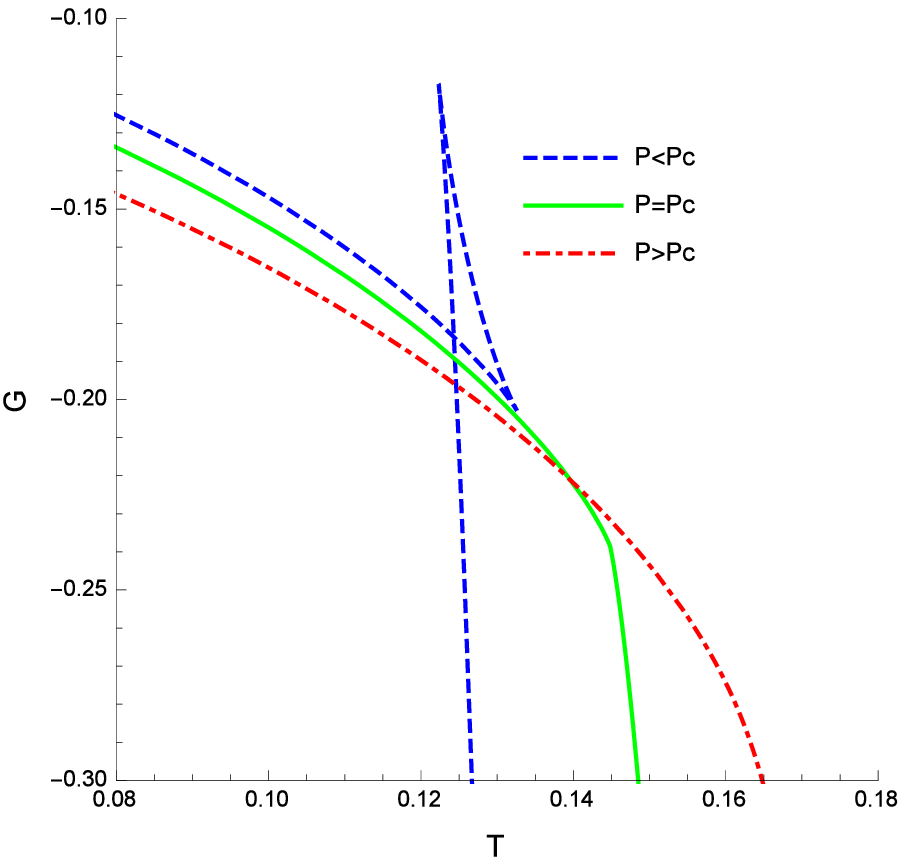}\label{Fig3b}}
	\caption{Gibbs diagrams of Lovelock black holes for $\hat{\eta}_{3}<0$ and $\kappa=0$}
	\label{Fig3}
\end{figure}

One of the most important items that helps us to determine phase transition
of a system refers to study its thermodynamic potential. Gibbs free energy generally
is computed from the Euclidean action with appropriate
boundary term \cite{Dayyanipld} while the
lowest Gibbs free energy is associated with global stable state. In the canonical
ensemble and extended phase space, thermodynamic potential closely
associates with the Gibbs free energy $G=M-TS$. As is well known,
to have a physical behavior, the second order derivative of Gibbs energy with
respect to the temperature should be negative to have a positive heat capacity.
Zeroth order phase transition occurs in the system when
Gibbs energy is discontinuous. This behavior was formerly observed in superfluidity and
superconductivity \cite{maslov}. Any discontinuity in fist (second) order derivatives
of Gibbs energy leads to a first (second) order phase transition in the system.
We calculate the Gibbs free energy of the black hole to
elaborate the phase transition of the system as below,
\begin{eqnarray}
G &=&G(P,T)=-\frac{Pr_{h}^{d-1}[5\alpha ^{2}(d-2)\eta _{2}+(d-6)r_{h}^{4}]}{%
(d-6)(d-2)(d-1)\left( \alpha ^{2}\eta _{2}+r_{h}^{4}\right) }  \notag \\
&&+\frac{\eta _{3}r_{h}^{d-7}[\alpha ^{4}(d-2)\left( d^{2}-3d+2\right) \eta
_{2}+5\alpha ^{2}(d-6)\left( d^{2}-3d+2\right) r_{h}^{4})}{48\pi
(d-6)(d-2)(d-1)\left( \alpha ^{2}\eta _{2}+r_{h}^{4}\right) }  \notag \\
&&+\frac{r_{h}^{d-7}\left( 9\alpha (d-6)\left( d^{2}-3d+2\right) \eta
_{2}r_{h}^{6}-3\alpha ^{3}(d-2)\left( d^{2}-3d+2\right) \eta
_{2}^{2}r_{h}^{2}\right) }{48\pi (d-6)(d-2)(d-1)\left( \alpha ^{2}\eta
_{2}+r_{h}^{4}\right) }  \label{G0}
\end{eqnarray}%
where $r_{h}$ should be understood as a function of pressure and temperature
via the equation of state.
The Gibbs free energy corresponding to Fig. \ref{Fig1} is depicted in Fig. \ref{Fig3}.
One can note that for negative $\eta _{3}$,
the Gibbs energy has a smooth behavior as a function of $T$, for $P>P_{c}$
whereas for $P<P_{c}$  it exhibits the small/large black hole first order phase transition
and a usual swallowtail shape as we expect
which is the characteristic of van der Waals fluid.

\begin{figure}[tbp]
	\centering
	\subfigure[$\alpha=1$, $\hat{\eta}_{2}=2$, $\hat{\eta}_{3}=5$ and
	$d=8$. One has two critical points with positive pressure while
	only the one with $P=P_{c2}$ corresponds to the first order phase
	transition between small and large black holes. The other
	does not globally minimize the Gibbs free energy and hence
	is unphysical.]{\includegraphics[scale=0.8]{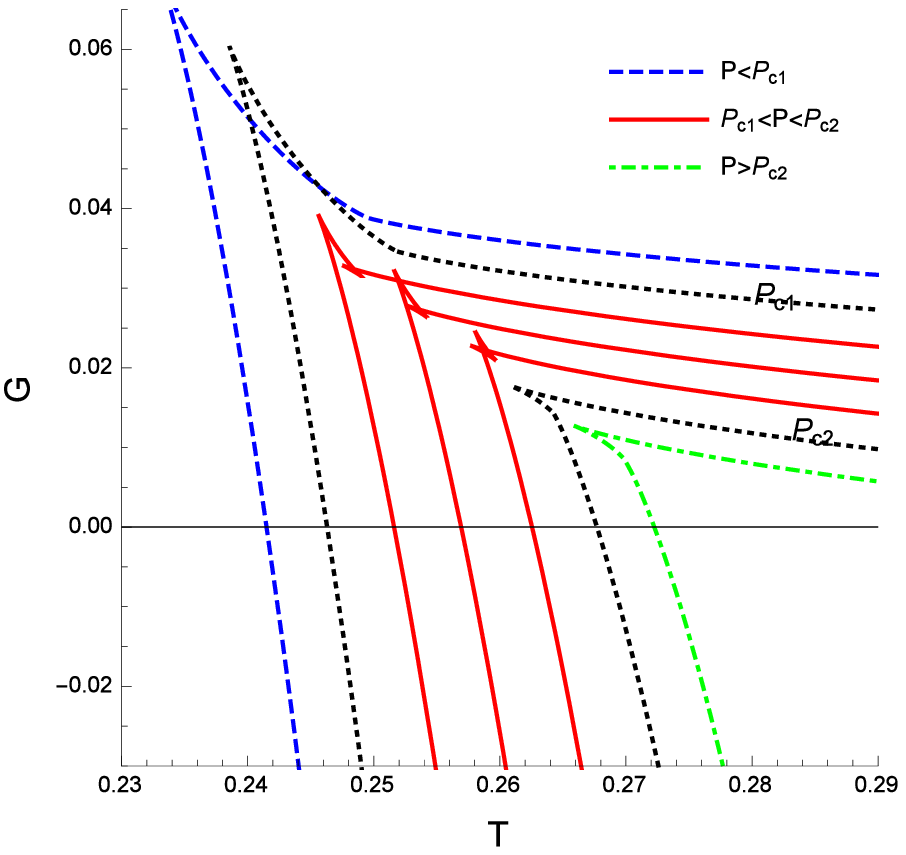}\label{Fig50a}} \hspace*{.2cm}
	\subfigure[$\alpha=1$, $\hat{\eta}_{2}=2$, $\hat{\eta}_{3}=2$ and
	$d=8$. There are two critical points with positive and negative pressure. As one expects
	only  $T=T_{c2}$ with positive pressure corresponds to the first order phase
	transition. The other critical point ($T=T_{c1}$) is unphysical]{\includegraphics[scale=0.8]{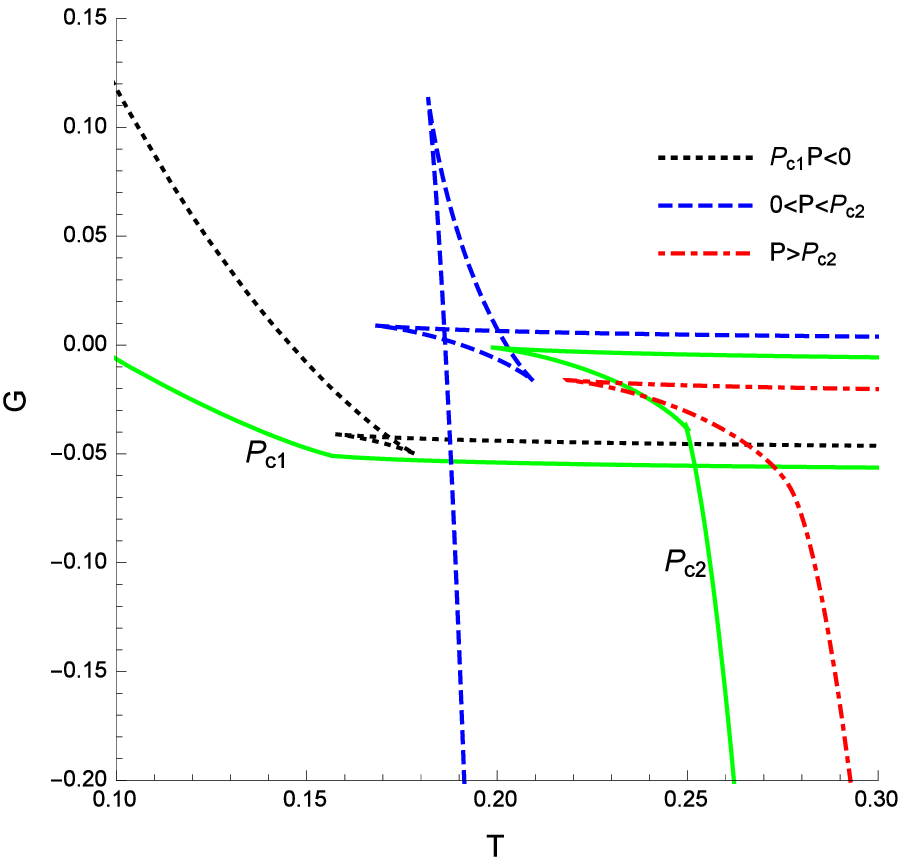}\label{Fig50b}}
	\caption{Gibbs diagrams of Lovelock black holes for $\hat{\eta}_{3}>0$ and $\kappa=0$. The curves are shifted and rescaled for more clarity.}
	\label{Fig50}
\end{figure}

The corresponding Gibbs diagram to Fig. \ref{Fig40} with two critical points
is displayed in Fig. \ref{Fig50}. In Fig. \ref{Fig50a} when $P \leq P_{c1} $ the lower (upper) branch
is thermodynamically stable(unstable). There is only one physical branch and Gibbs energy
shows no phase transition in the system. However, a first order phase transition may happen
in the range of $P_{c1}<P<P_{c2}$ as shown with solid red line in  Fig. \ref{Fig50a}.
In Fig. \ref{Fig50b} we can see a first order phase
transition similar to Van der Waals fluid in the range $0<P<P_{c2}$ and thus the
second critical point (with $P=P_{c2}$) is physical.
The critical point with negative pressure does not globally minimize the Gibbs energy
and hence is not physical.
Although the equation of state leads to one or two critical point in this case,
investigating the Gibbs diagrams represents only one physical critical point.

\subsection{Critical exponents}

\label{exponent0}

Critical exponent characterizes the behavior of physical quantities in the
close vicinity of the critical point. We proceed to calculate the
critical exponents $\alpha ^{\prime }$, $\beta ^{\prime }$, $\gamma ^{\prime
}$ and $\delta ^{\prime }$ for the phase transition of a $d$-dimensional
Lovelock black hole. In order to calculate the first critical exponent\ $%
\alpha ^{\prime }$, we consider the entropy $S$ given by Eq. (\ref{Entro}) as a
function of $T$ and $v$. Making use of Eq.(\ref{ESV}) we have
\begin{equation}
S=S\left( T,v\right) =4^{1-d}(d-2)^{d-2}v^{d-2}+\frac{\alpha
^{2}4^{5-d}(d-2)^{d-5}v^{d-6}\hat{\eta} _{2}}{d-6}
\end{equation}%
It is clear that entropy does not depend on the temperature in this
relation and hence $C_{V}=0$. This indicates that relative critical exponent will
be zero
\begin{equation}
C_{V}\propto \left( \frac{T}{T_{c}}-1\right) ^{\alpha ^{\prime }}\Rightarrow
\alpha ^{\prime }=0.
\end{equation}

To obtain the other exponents, we define the reduced thermodynamic variables as
\begin{equation*}
p\equiv \frac{P}{P_{c}},\quad \nu \equiv \frac{v}{v_{c}},\quad \tau \equiv
\frac{T}{T_{c}},
\end{equation*}%
and expansion parameters as
\begin{equation}
t=\tau -1,\quad \omega =\nu -1=\dfrac{v}{v_{c}}-1.  \label{29}
\end{equation}
Then we can make Taylor expansion for the equation of state Eq.(\ref{Pk0})
as
\begin{equation}
p=1+At-Bt\omega -C\omega ^{3}+O\left( t\omega ^{2},\omega ^{4}\right) ,
\label{ptw}
\end{equation}
where $A$, $B$ and $C$ are constants depending on $d$, $\alpha$, $\hat{\eta}_{2}$ and $%
\hat{\eta}_{3}$.

Denoting the volume of small and large black holes by $\omega _{s}$ and $%
\omega _{l}$, respectively, differentiating Eq. (\ref{ptw}) with respect to
$\omega $\ at a fixed $t<0$, and applying the Maxwell's equal area law \cite%
{Spall} one obtains
\begin{eqnarray}
p &=&1+At-Bt\omega _{l}-C\omega _{l}^{3}=1+At-Bt\omega _{s}-C\omega _{s}^{3}
\notag \\
0 &=&-P_{c}\int_{\omega _{l}}^{\omega _{s}}\omega \left( Bt+3C\omega
^{2}\right) d\omega ,
\end{eqnarray}
which leads to the unique non-trivial solution
\begin{equation}
\omega _{l}=-\omega _{s}=\sqrt{-\frac{Bt}{C}}.  \label{oml}
\end{equation}
Thus, the exponent $\beta ^{\prime }$, which describes the behaviour of the
order parameter $\eta =v_{c}\left( \omega _{l}-\omega _{s}\right) $ on a
given isotherm, may be calculated through the use of Eq. (\ref{oml}) as:
\begin{equation}
\eta =2v_{c}\omega _{l}=2\sqrt{-\frac{Bt}{C}}\quad \Longrightarrow \quad
\beta ^{\prime }=\frac{1}{2}.
\end{equation}
To calculate the exponent $\gamma ^{\prime }$, we may determine the behavior
of the isothermal compressibility near the critical point
\begin{equation*}
\kappa _{T}=-\frac{1}{V}\frac{\partial V}{\partial P}\Big|_{T}\varpropto
\left\vert t\right\vert ^{-\gamma ^{\prime }}.
\end{equation*}
Since $dv/d\omega =v_{c}$, the isothermal compressibility near the critical
point reduces to
\begin{equation}
\kappa _{T}=-\frac{1}{V}\frac{\partial V}{\partial P}\Big|_{T}\varpropto
\frac{V_{c}}{BP_{c}t},
\end{equation}
which shows that $\gamma ^{\prime }=1$. Finally the `shape` of the critical
isotherm\ $t=0$ is given by (\ref{ptw})
\begin{equation}
p-1=-C\omega ^{3},
\end{equation}
which indicates that $\delta ^{\prime }=3$.

The critical exponents associated with this type of Lovelock black holes are
independent of metric parameters and the dimension of the spacetime. This is
consistent with the results of mean field theory that believe
the critical exponents are universal and do not depend on the details of the
physical system.

\section{Critical behavior of Lovelock spherical black holes with $\protect\kappa =1$}

\label{criticalitys}

When the topology of the black hole horizon is spherical the equation of
state is in the form

\begin{equation}
P=\frac{T}{v}-\frac{d-3}{\pi (d-2)v^{2}}+\frac{32\alpha T}{(d-2)^{2}v^{3}}-%
\frac{16\alpha (d-5)\left[\hat{\eta} _{2}+1\right] }{\pi (d-2)^{3}v^{4}}+\frac{%
256\alpha ^{2}\left[\hat{\eta}_{2}+1\right] T}{(d-2)^{4}v^{5}}-\frac{256\alpha
^{2}(d-7)\left[ 3\hat{\eta}_{2}+\hat{\eta}_{3}+1\right] }{3\pi (d-2)^{5}v^{6}}
\label{Psph}
\end{equation}
In different types of black holes, first order phase transition occurs
for black holes which have spherically symmetric horizon. Therefor it is important to investigate
this case $(\kappa =1)$ and compare our results with the other types of
black holes.

It is shown in \cite{Xu}, that for third order Lovelock black holes with $%
\kappa =1$ and constant curvature horizon there exist two critical points
for $8\leq d\leq 11$ and no critical point exists for $d>11$.

For our solution with nonconstant curvature horizon, Eqs. (\ref{CritEq}) could
be simplified as a polynomial of degree 4 as:
\begin{equation}
v^{4}+bv^{3}+cv^{2}+dv+e=0  \label{Quartic Eq}
\end{equation}%
with%
\begin{eqnarray}
b &=&96\frac{\hat{\eta}_{2}(d-5)-2}{d-3}  \notag \\
c &=&256\frac{5(d-7)\hat{\eta} _{3}+12(d-10)\hat{\eta} _{2}+2(d-25)}{d-3}  \notag \\
d &=&8192\frac{9(d-7)\hat{\eta} _{3}-5(d-5)\hat{\eta} _{2}^{2}+(17d-139)\hat{\eta}
_{2}+2(2d-19)}{d-3}  \notag \\
e &=&327680(d-7)\frac{[(\hat{\eta} _{2}+1)\hat{\eta} _{3}+1]+3\hat{\eta} _{2}^{2}+4\hat{\eta} _{2}}{%
d-3}.  \label{Parameters of QEq}
\end{eqnarray}

A well-known calculation yields%
\begin{eqnarray}
\Delta
&=&b^{2}c^{2}d^{2}-4b^{2}c^{3}e-4b^{3}d^{3}+18b^{3}cde-27b^{4}e^{2}-4c^{3}d^{2} \notag \\
&&+16c^{4}e+18bcd^{3}-80bc^{2}de-6b^{2}d^{2}e+144b^{2}ce^{2}\notag \\
&&-27d^{4}+144cd^{2}e-128c^{2}e^{2}-192bde^{2}+256e^{3}.
\label{Delta}
\end{eqnarray}
For the equation (\ref{Quartic Eq}) to have solution, $\Delta $ should be positive,
which again makes a constraint on the parameter $\hat{\eta} _{3}$ relating to $%
\hat{\eta} _{2},$ $\alpha $ and $d.$ On the other hand, a look at the relation (\ref{Psph})
reveals that the dominant term is the last one, which is positive for $\hat{\eta}_3 \leq -(3\hat{\eta}_{2}+1)$.
If $\hat{\eta}_3$ is chosen in such a way that this inequality and also $\Delta>0$ hold,
the pressure tends to $+\infty$ as $v\rightarrow0$ and
Eq. (\ref{Quartic Eq}) has one real root. Thus the system demonstrates
Van der Waals behavior.
For the values of $\hat{\eta}_{3}$ satisfying $\hat{\eta}_{3}>-(3\hat{\eta}_{2}+1)$, and the constraint $\Delta>0$,
the pressure tends to $-\infty$ as $v\rightarrow0$ and we may have up to two critical points. Various physical
situations are summarized in table
\ref{table1} for three values of $d$.

\begin{table}[!h]
\tabcolsep 0pt
\caption{Critical values in different dimensions for $\kappa=1$}
\vspace*{-12pt}
\begin{center}
\def\temptablewidth{0.7\textwidth}
{\rule{\temptablewidth}{1pt}}

\begin{tabular}{llllllllll}
$\ \ \ d\ \ \ $ & $\ \ \alpha \ \ $ & $\ \ \eta _{2}\ \ $ & $\ \ \ \eta
_{3}\ \ \ \ $ & $\ \ \ \ \ v_{c1}\ \ \ \ \ $ & $\ \ \ \ \ \ T_{c1}\ \ \ \ \ $
& $\ \ \ \ \ P_{c1}\ \ \ \ \ $ & $\ \ \ \ \ v_{c2}\ \ \ \ $ & $\ \ \ \ \
T_{c2}\ \ \ $ & $\ \ \ \ \ P_{c2}\ \ $ \\
$\ \ \ 8$ & $\ \ 1$ & $\ \ 1$ & $\ \ 0.8$ & $\ \ 0.5632$ & $\ \ 0.2027$ & $%
\ \ \ 0.0133$ & $\ \ 1.1376$ & $\ \ 0.2131$ & $\ \ 0.0630$ \\
$\ \ \ 8$ & $\ \ 1$ & $\ \ 1$ &  $-0.5$ & $\ \ 0.4369$ & $\ \ 0.1823$ & $\
-0.1481$ & $\ \ 1.2124$ & $\ \ 0.2101$ & $\ \ 0.0599$ \\
$\ \ \ 8$ & $\ \ 1$ & $\ \ 1$ & $\ -5$ & $\ \ 1.3750$ & $\ \ 0.2023$ & $\ \
\ 0.0531$ & \ \ \ \ --- & \ \ \ \ --- & \ \ \ \ --- \\
$\ \ \ 9$ &  $\ \ 1$ & $\ \ 1$ &  $-0.5$ & $\ \ 0.5914$ & $\ \ 0.2716$ & $\ \
\ 0.1007$ & $\ \ 0.8118$ & $\ \ 0.2728$ & $\ \ 0.1067$ \\
$\ \ \ 9$ & $\ \ 1$ & $\ \ 1$ & $\ -3$ & $\ \ 0.2201$ & $\ \ 0.1707$ & $\ \
-3.0175$ & $\ \ 1.050$ & $\ \ 0.8346$ & $\ \ 0.0879$ \\
$\ \ \ 9$ & $\ \ 1$ & $\ \ 1$ & $\ -5$ & $\ \ 1.1433$ & $\ \ 0.2516$ & $\ \
\ 0.0805$ & \ \ \ \ --- & \ \ \ \ \ --- & \ \ \ \ --- \\
$\ \ 12$ & $\ \ 1$ & $\ \ 1$ & $\ -2$ & $\ \ 0.3700$ & $\ \ 0.4399$ & $\ \ \
0.2573$ & $\ \ 0.5193$ & $\ \ 0.4422$ & $\ \ 0.2793$ \\
$\ \ 12$ & $\ \ 1$ & $\ \ 1$ & $\ -3$ & $\ \ 0.1907$ & $\ \ 0.3460$ & $\ \
-2.0575$ & $\ \ 0.6479$ & $\ \ 0.4230$ & $\ \ 0.0234$ \\
$\ \ 12$ & $\ \ 1$ & $\ \ 1$ & $\ -5$ & $\ \ 0.7605$ & $\ \ 0.4001$ & $\ \ \
0.1960$ & \ \ \ \ --- & \ \ \ \ \ --- & \ \ \ \ ---%
\end{tabular}

       {\rule{\temptablewidth}{1pt}}
       \end{center}
       \label{table1}
       \end{table}

\begin{figure}
 	\centering \subfigure[$d=8, \alpha=1$, $\hat{\eta}_{2}=1$ and $\hat{\eta}_{3}=-5$.]{\includegraphics[scale=0.55]{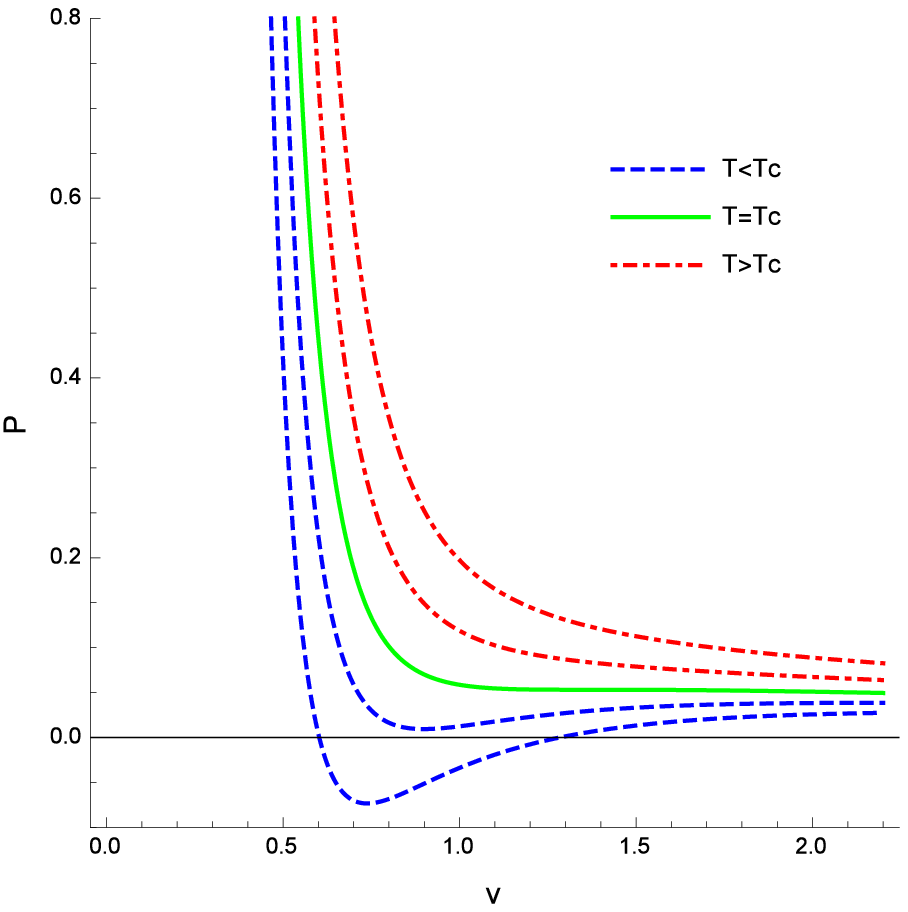}\label{Fig4a}}
 	\hspace*{.2cm}
   \centering \subfigure[$\alpha=1$, $\hat{\eta}_{2}=1$ and $\hat{\eta}_{3}=0.8$.]{\includegraphics[scale=0.7]{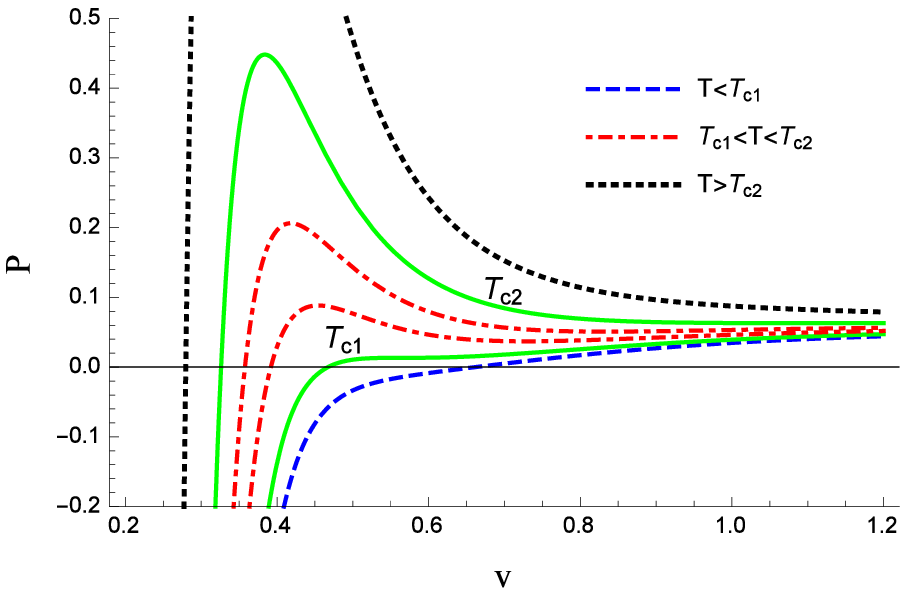}\label{Fig4b}}
   \hspace*{.2cm}
 	\centering \subfigure[$d=8, \alpha=1$, $\hat{\eta}_{2}=1$ and $\hat{\eta}_{3}=-0.5$.]{\includegraphics[scale=0.7]{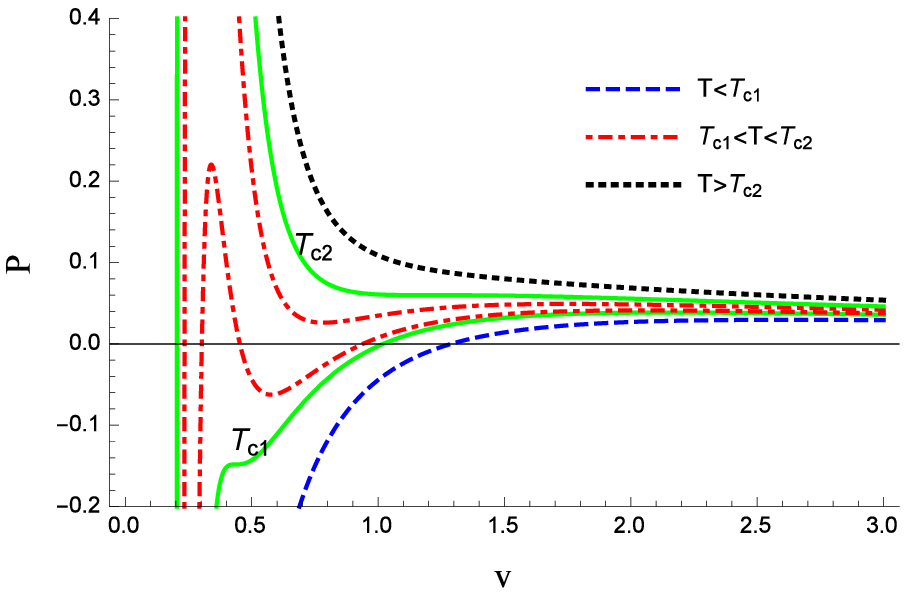}\label{Fig4c}}
 	\hspace*{.2cm}
  \caption{$P-v$
 		Diagram of Lovelock black holes with spherical horizon, $\kappa=1$, in $d=8$ which shows a Van der Waals behavior (a), and
  the possibility of reentrant phase
 transition (b) and (c).}\label{Fig4}
 \end{figure}

The corresponding $P-v$ diagrams
for diverse choices of $\eta_{3}$ are depicted in Fig \ref{Fig4}. The diagrams are the same
in any dimension $d$. So without loss of generality we plot them for $d=8$.
The interesting point that one should note is that despite of the case for the black holes with constant curvature horizon,
there exists criticality for $d\geq11$.
As one can see, for some choices of $\hat{\eta}_{3}$ two critical points are present, one with negative
(unphysical) and the other with positive pressure (\ref{Fig4c}). But one can find values for free parameters
for which two critical points with positive pressure exist as is seen in Fig. \ref{Fig4b}. This behavior is
reminiscent of the interesting reentrant phase transition. So we go through the Gibbs plot for these cases.
The Gibbs free energy obeys the following thermodynamic relation for any value of $\kappa$
\begin{eqnarray}
G &=&\frac{Pr^{d-1}}{d-1}+\frac{(d-2)kr^{d-3}}{16\pi }+\frac{\alpha
(d-2)r^{d-5}\left(\hat{\eta}_{2}+k^{2}\right) }{16\pi }+\frac{\alpha
^{2}(d-2)r^{d-7}\left( 3\hat{\eta}_{2}k+\hat{\eta} _{3}+k^{3}\right) }{48\pi }  \notag
\\
&&-\frac{r^{d-7}\left( \alpha ^{2}(d-4)(d-2)\left(\hat{\eta} _{2}+k^{2}\right)
+2\alpha (d-6)(d-2)kr^{2}+(d-6)(d-4)r^{4}\right) }{48\pi
(d-6)(d-4)(d-2)\left( \alpha ^{2}\hat{\eta} _{2}+\left(\alpha k+r^{2}\right)
^{2}\right) }  \notag \\
&&\times \left\{ \alpha ^{2}(d-7)(d-2)\left( 3\hat{\eta} _{2}k+\hat{\eta}
_{3}+k^{3}\right) +3\alpha (d-5)(d-2)r^{2}\left(\hat{\eta} _{2}+k^{2}\right)
+3(d-3)(d-2)kr^{4}+48\pi Pr^{6}\right\}   \label{G,k}
\end{eqnarray}

\begin{figure}
 	\centering \subfigure[The lower (upper) branch  is thermodinamically stable(unstable). There is only one physical branch and no phase transition occurs in the system.
 	]{\includegraphics[scale=0.47
 		]{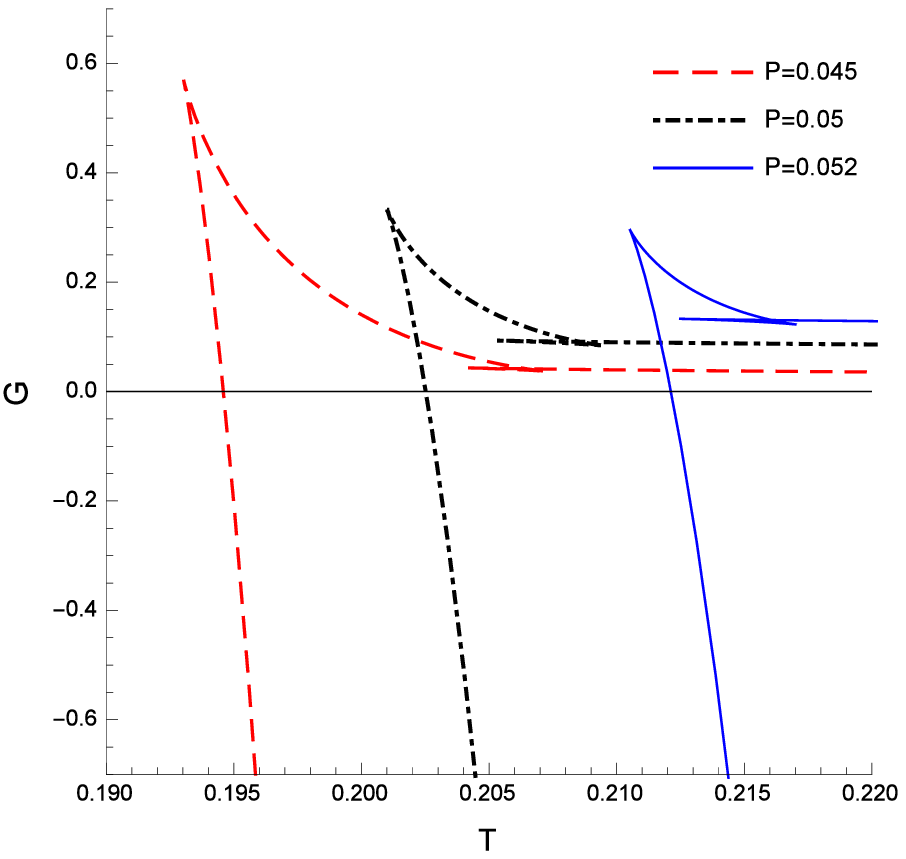}\label{Fig41a}}
 	\hspace*{.15cm} \subfigure[A first order phase transition occurs in $T_1$ and a zeroth order in $T_0$. In addition a finite jump in Gibbs diagram leads to
 	a zeroth order phase transition between small and  large (intermediate) black hole.]{\includegraphics[scale=0.32
 	 	]{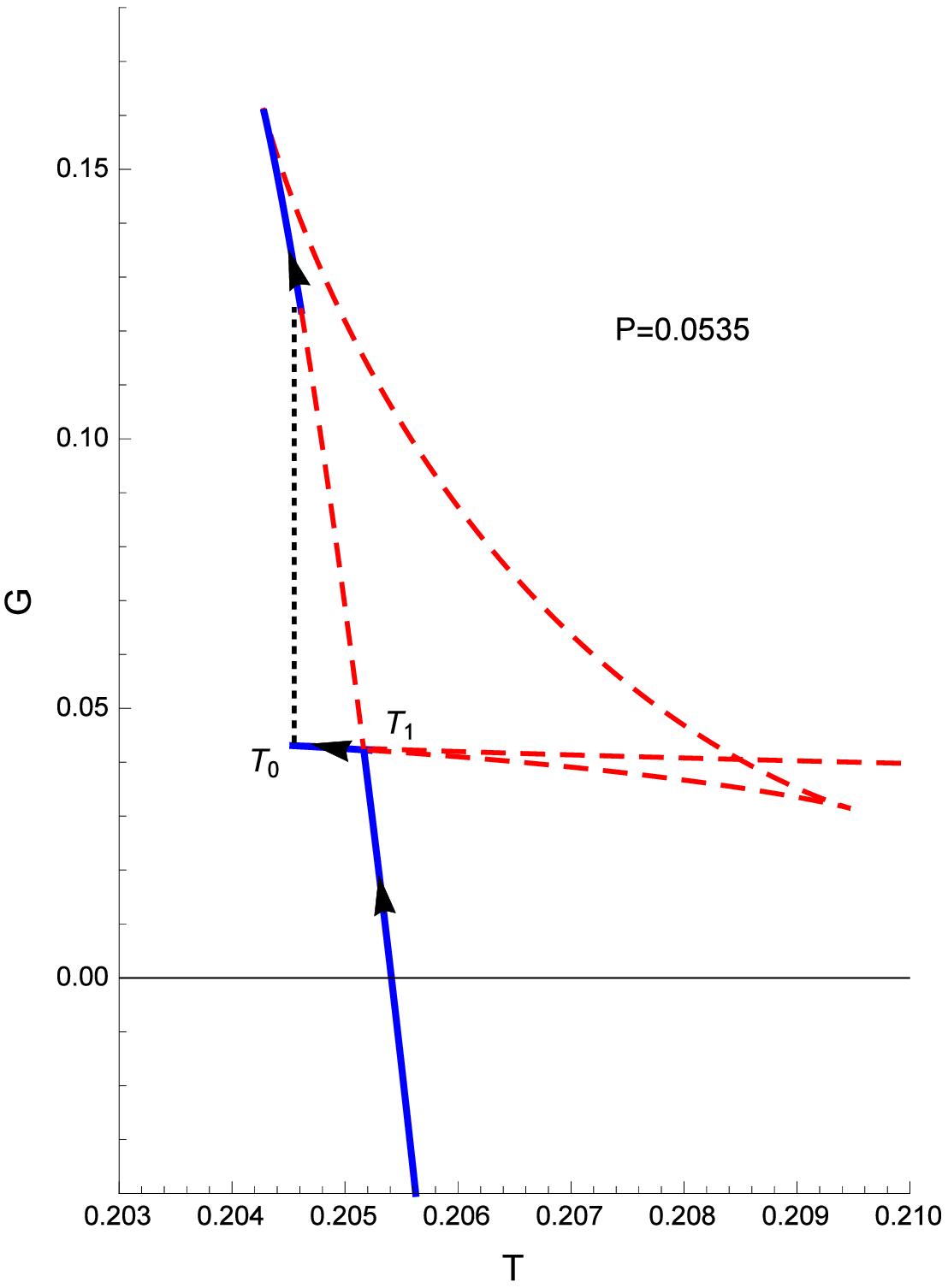}\label{Fig41b}}
 	\hspace*{.15cm} \subfigure[The first order phase transition finishes in the critical point $P=P_{c2}$. For $P>P_{c2}$, we see  only one stable (lower) branc with no phase transition. ]{\includegraphics[scale=0.4
 		]{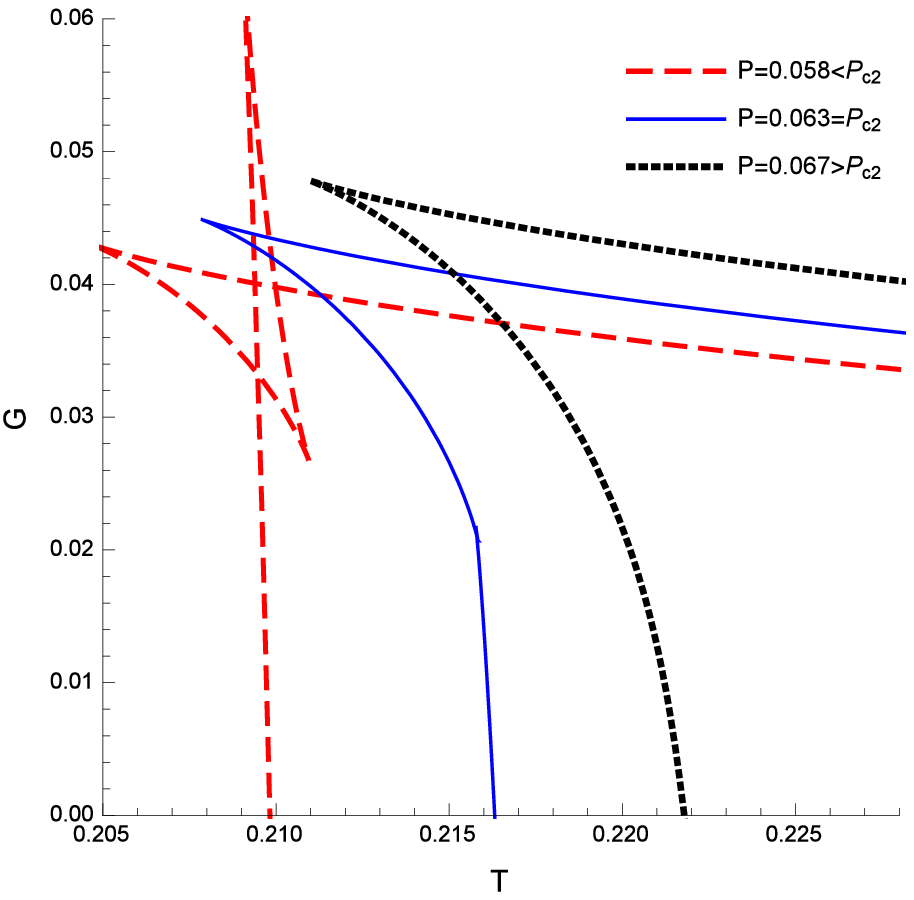}\label{Fig41c}}
 	\caption{Gibbs diagram with $d=8$, $\alpha=1$, $\hat{\eta}_2=1$ and $\hat{\eta}_3=0.8$ .}\label{Fig41}
 \end{figure}

 The behavior of Gibbs free energy for a positive value of $\hat{\eta}_{3}$ is depicted in Fig. \ref{Fig41}. Two positive critical pressures are $T_{c1}$ and $T_{c2}$. Looking at Fig. \ref{Fig41a} we observe that between $P_{c1}=0.013$ and $P=0.052<P_{c2}$, the lower branch is globally stable and therefore no phase transition could happen.

  By increasing the pressure, zeroth and first order phase transition take place between $P=0.053$ and $P=0.054$ (See Fig. \ref{Fig41b}). The first order phase
  transition occurs between small black hole (SBH) and large black hole (LBH).
  As well as the first order phase transition, a finite jump in Gibbs free energy is seen in the mentioned region which leads to a
  zeroth order phase transition between SBH and  LBH (or intermediate black hole). Interestingly enough,
  in the above certain range of pressure, a reentrant phase transition occurs.
  Heretofore, zeroth order and reentrant phase transition has been detected in charged dilaton
  \cite{Dehy1} and  Born-Infeld \cite{Kubiz,Dehy2} black holes.
   A reentrant phase transition  is a Combination of
   two (or more) phase transition, in such a way that the initial and final phase of system are macroscopically
   the same.

  By increasing the pressure, in the region $0.054<P \leq P_{c2}=0.063$ the zeroth order phase transition disappears and only first order phase transition could bee seen. Finally, the first order phase transition finishes at second critical point $P=P_{c2}=0.063$
 as is seen in Fig. \ref{Fig41c}. For $P>P_{c2}$, the lower branch is the unique globally stable branch and we do not expect any  phase transition.

 \begin{figure}[h]
	\centering \subfigure[A zeroth order phase transition occurs in $T_0$.
	The only stable branch is lower one which is showed by solid blue line. The diagram showes a finite jump between stable and unstable branch and physically can not exist.]{\includegraphics[scale=0.5
		]{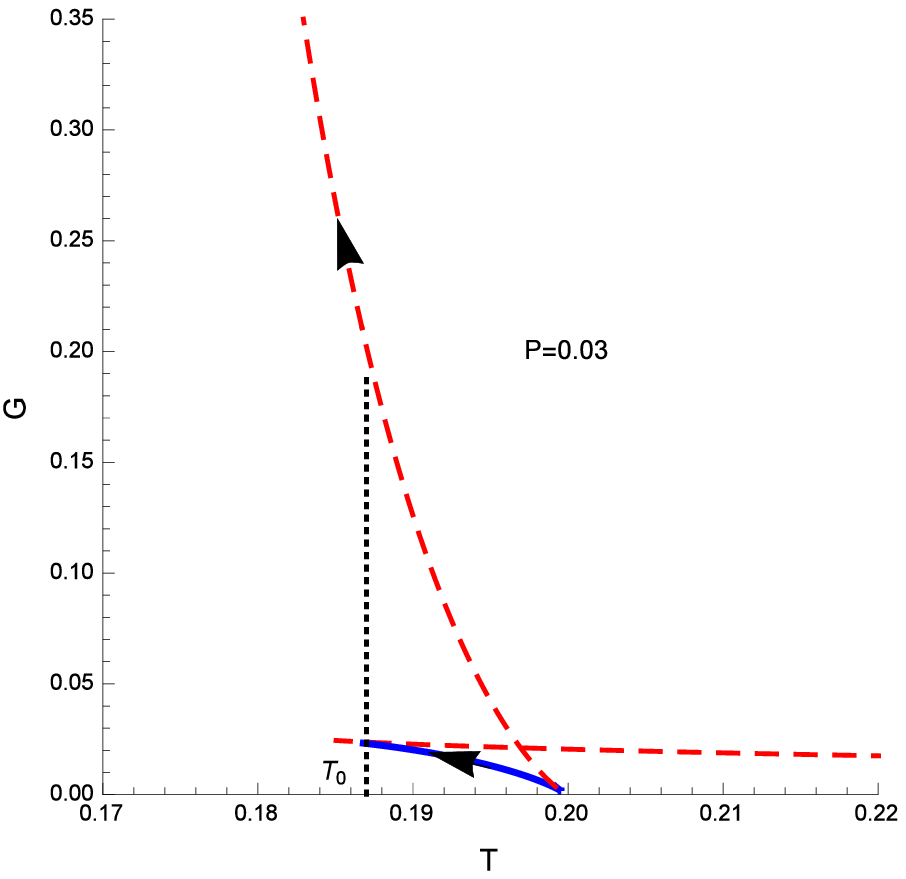}\label{Fig31a}}
	\hspace*{.15cm} \subfigure[The first order phase transition occurs in $T_1$
	and a zeroth order in $T_0$. A reentrant phase transition occurs between  LBH /SBH /LBH.
	]{\includegraphics[scale=0.3
		]{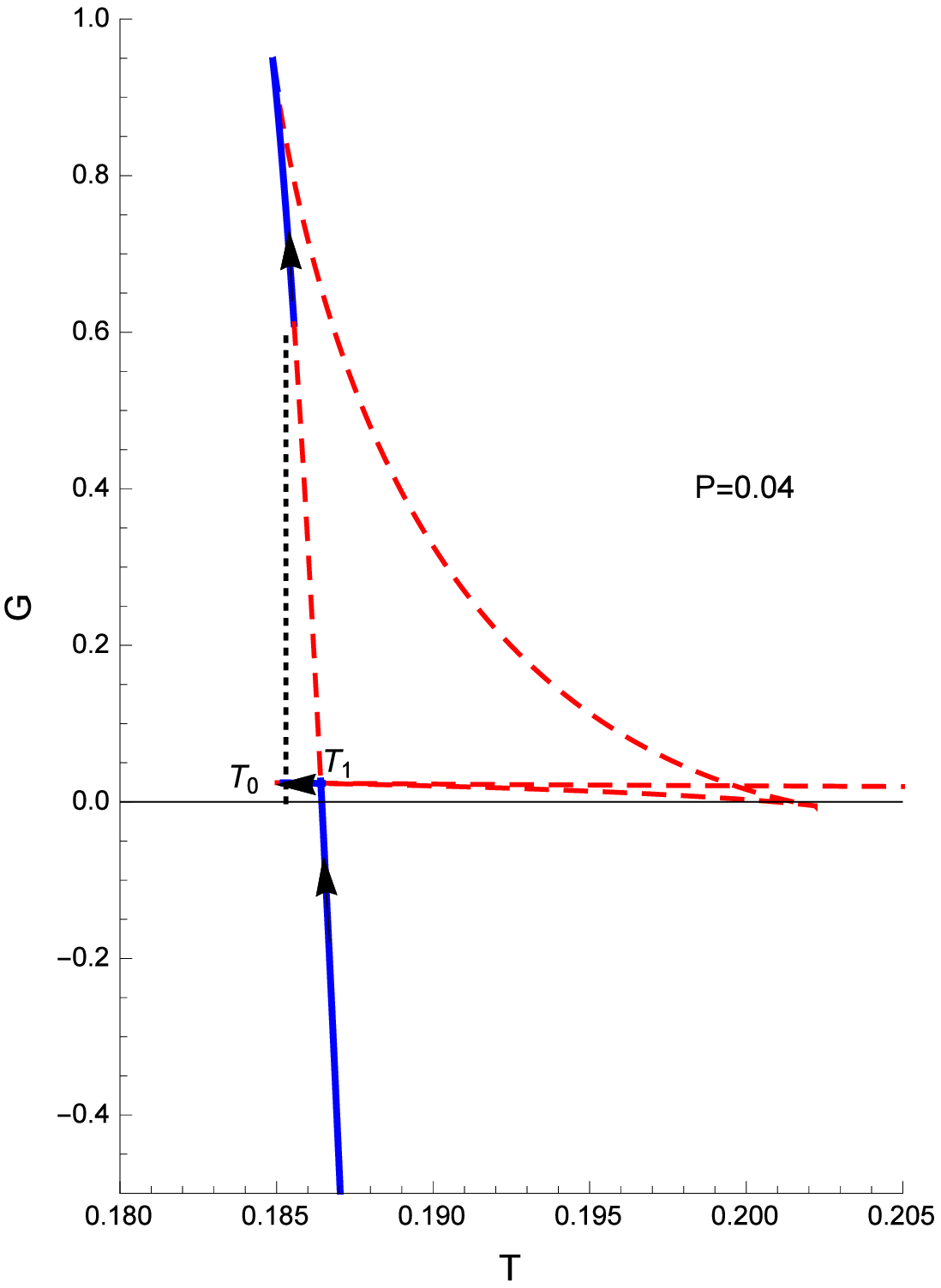}\label{Fig31b}}
	\hspace*{.15cm} \subfigure[The first order phase transition finished in  the critical point $P=P_{c2}$. There is only one stable (lower) branch For $P>P_{c2}$ and so no phase transition occurs.]{\includegraphics[scale=0.4
		]{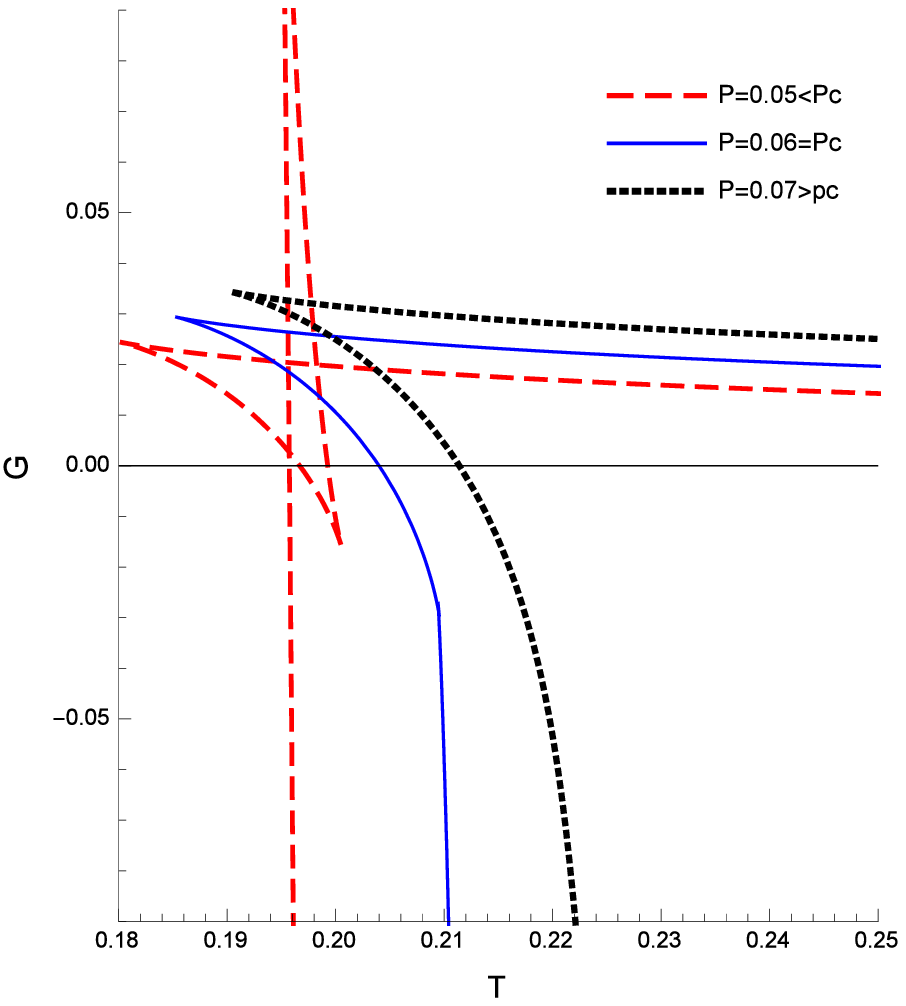}\label{Fig31c}}
	\caption{Gibbs diagram with $d=8$, $\alpha=1$, $\hat{\eta}_2=1$ and $\hat{\eta}_3=-0.5$.}\label{Fig31}
\end{figure}

As one can see in Fig. \ref{Fig4c}, there are two critical points with positive and negative pressure. Here we would like to study the associated Gibbs free energy in Fig. \ref{Fig31}. In the region $0<P<0.032$, there exist a discontinuity in the Gibbs diagram (See Fig. \ref{Fig31a}). In this figure, the blue solid line shows reasonable branch of Gibbs energy and dashed lines show unstable parts of Gibbs energy with negative heat capacity. One may observe two physical phase transitions for $P=0.04$ in Fig.\ref{Fig31b}. A first order phase transition occurs at $T_1$, where the Gibbs diagram is continuous but its derivative is not. Also, there is a finite gap in Gibbs diagram in $T_0$ and so a zeroth order phase transition occurs in this point. Fig.\ref{Fig31b} represents a reentrant phast transition between LBH/ SBH/ LBH . More increasing of the pressure leads to the disappearance of zeroth order phase transition. For example in $P=0.055$, there is no zeroth order phase transition. Fig. \ref{Fig31c} shows three different states of Gibbs energy around the second critical pressure $(P_{c2}=0.06)$. For $P=0.055<P_{c2}$, one sees a first order phase transition and for $P=0.07>P_{c2}$, no phase transition happens in the system. While $T=T_{c2}$ is the critical point for which second order phase transition occurs, because of the existence of only one stable branch for $P>P_{c2}$, no phase transition may occur.

 \begin{figure}
 	\epsfxsize=8cm \centerline{\epsffile{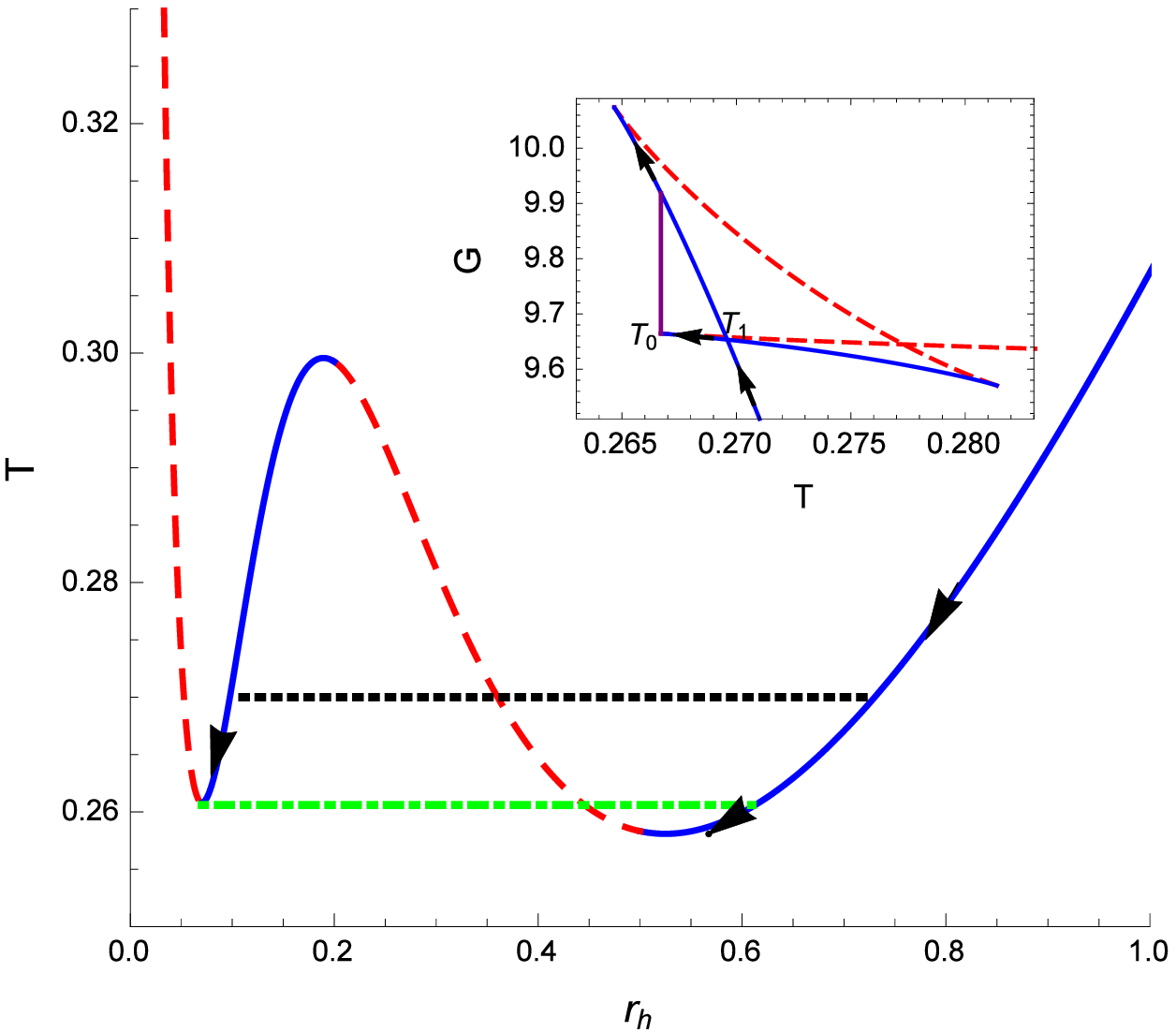}}
 	\caption{ The  schematic behavior of isobar $T$-$r_h$ diagram and the
 		corresponding $G-T$
 		 curve (inset) of Lovelock black hole. As temperature decreases
 		the black hole follows direction of arrows. The first and zeroth
 		order phase transition are identified by black dotted and green dot-dashed curve,
 		respectively.
 		A large/small/large(intermediate) corresponds
 		to a reentrant phase transition. The positive (negative) sign
 		of heat capacity
 		is displayed by the blue solid (dash red) line. We have changed the size of diagram and shifted for more clarity but this diagram can represent the behavior of Figs. \ref{Fig41b} and \ref{Fig31b}.}
 	\label{Fig51}
 \end{figure}

Now, we go through the study of the behavior of isobar $T-r_h$ diagram and the
corresponding $G-T$
curve (inset) for Lovelock black hole with nonconstant curvature horizon. As it is illustrated in Fig. \ref{Fig51} by decreasing the radius of horizon in the $G-T$ plane (the inset in Fig. \ref{Fig51}), black hole follows the lower solid blue branch until
it reaches $T_1$ and changes the direction to switch to left solid blue curve in $G-T$ plane with a first order LBH/SBH
phase transition. This is identified by black dotted line in $T-r_h$ plane. In the case of more decreasing of $r_h$, the system experiences zeroth order phase transition between small and large black hole at $T_0$ by a finite jump ( inset in Fig. \ref{Fig51}) which  is shown by dot-dashed green line in $T-r_h$ diagram in Fig. \ref{Fig51}. Eventually, the black hole tracks the blue solid line to the end. We should mention that we have shifted the diagram for more clarity but this diagram can represent the behavior of Figs. \ref{Fig41b} and \ref{Fig31b}.

  The same as what we observed in the case of Ricci flat black holes with $\kappa=0$ in Sec. \ref{criticalityf}, the entropy does not depend on the temperature and so the first exponent $\alpha^\prime$ equals zero for $\kappa=1$.
  Following the approach discussed in Sec.\ref{exponent0}, we can write the reduced equation of state as
  \begin{equation}
  p=1+At-Bt\omega -C\omega ^{3}+O\left( t\omega ^{2},\omega ^{4}\right).
   \end{equation}
  Therefore, it is easy to show that the critical exponents read
   \begin{equation}
     \beta ^{\prime }=\frac{1}{2},\quad \gamma ^{\prime }=1, \quad \delta ^{\prime }=3
   \end{equation}

\section{Critical behavior of Lovelock Hyperbolic black holes with $\protect\kappa =-1$}

\label{criticalityh}

When the topology of the black hole horizon is hyperbolic, the equation of
state reads
\begin{equation}
P=\frac{T}{v}+\frac{d-3}{\pi (d-2)v^{2}}-\frac{32\alpha T}{(d-2)^{2}v^{3}}-%
\frac{16\alpha (d-5)(\hat{\eta}_2+1)}{\pi (d-2)^{3}v^{4}}+\frac{256\alpha
^{2}(\hat{\eta}_2+1)T}{(d-2)^{4}v^{5}}-\frac{256\alpha ^{2}(d-7)(-3\hat{\eta}_2+\hat{\eta}
_3-1)}{3\pi (d-2)^{5}v^{6}}.  \label{Ph}
\end{equation}
In addition, the Gibbs free energy of the black hole can be calculated from Eq.(\ref{G,k}) by substituting $\kappa =-1$.

For the black hole with hyperbolic horizon in different theories of electrodynamics and general relativity,
first order phase transition is rarely seen. One critical point exists for second and third order Lovelock hyperbolic black holes with
constant curvature horizons \cite{Xu}. So, it is interesting to search for such a phase transition
for Lovelock black holes with nonconstant curvature horizon when $\kappa =-1$.

Similar to the case with  $\kappa =1$ which we discussed in Sec. \ref{criticalitys},
the equation of state, (Eqs. \ref{CritEq}) leads to a polynomial of degree 4 and numeric calculations
show that depending on the values of the parameters $\alpha$, $\hat{\eta}_{2}$, and $\hat{\eta}_{3}$
there may exist one or two physical critical points. In addition, we find some values for the free parameters
that results in three critical points,
with two of them having positive pressure and one with negative pressure. We could not find any case with
three positive critical pressure.

\begin{table}[!h]
	\tabcolsep 0pt
    \caption{Critical values in different dimensions for $\kappa=-1$}
	\vspace*{-12pt}
	\begin{center}
		\def\temptablewidth{0.9\textwidth}
		{\rule{\temptablewidth}{1pt}}
		\begin{tabular*}{\temptablewidth}{@{\extracolsep{\fill}}cccccccccccccccccc}
			\hline
			\ \ \ $\ d$\ \ \ \ \  & \ \ \ $\ \alpha \ \ \ \ \ $ & \ \ \ $\ \hat{\eta} _{2}$\ \
			\ \ \  & $\ \ \ \ \ \hat{\eta} _{3}\ \ \ \ \ \ $ & \ \ \ \ $T_{c1}$\ \ \ \ \ \  &
			 \ \ \ \ $\ P_{c1}$\ \ \ \ \ \  & \ \ \ \ $T_{c2}$\ \ \ \ \ \  & \ \ \ \ $\ P_{c2}$\ \ \ \ \ \  &
			\ \ \ \ $\ T_{c3}$\ \ \ \ \ \  &  \ \ \ \ $\ P_{c3}$\ \ \ \ \ \  \\ \hline
			8 & 1 & 3 & 1.1 &  \ 0.18162 &  \ 0.1770 & \ \ --- & \ --- & \ \ --- & \ --- \\ \hline
			10 & 1 & 10 & \ -3.5 &  0.3887 &  \ 0.2782 & \  --- & \   --- & \ \ --- & \ --- \\ \hline
			8 & 1 & \ 0.5 & \ -0.5 &  \ 0.0636 &  \ 0.2268 & \ 0.9091 & \ 0.5400 & \ \ --- & \ --- \\ \hline
			9 & 1 & \ 0.5 & \ 0.1 &  \ 0.0823 &  \ 0.3368 & \   0.9251 & \  0.38650 & \ \ --- & \ --- \\ \hline
			9 & 1 & \ 0.7 & \ 3.2 &  \ 0.0622 &  \ -306.37 & \ 0.8342 & \ 0.6184  & \ \ 1.2620 & \ 0.8414 \\
			\hline
		
		\end{tabular*}
		
		{\rule{\temptablewidth}{1pt}}
	\end{center}
	\label{table3}
\end{table}

\begin{figure}[tbp]
	\centering
	\subfigure[$P-v$ diagram]{\includegraphics[scale=0.6]{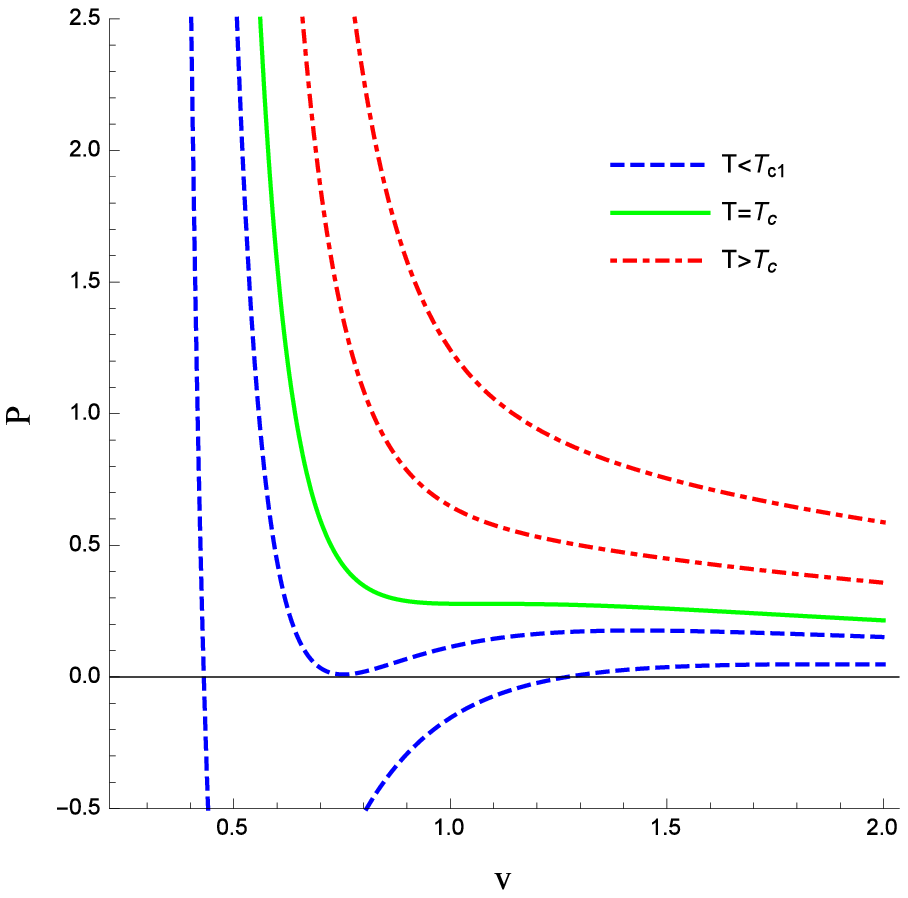}\label{Fig12b}} \hspace*{.2cm}
	\subfigure[Gibbs diagram.
	]{\includegraphics[scale=0.6
			]{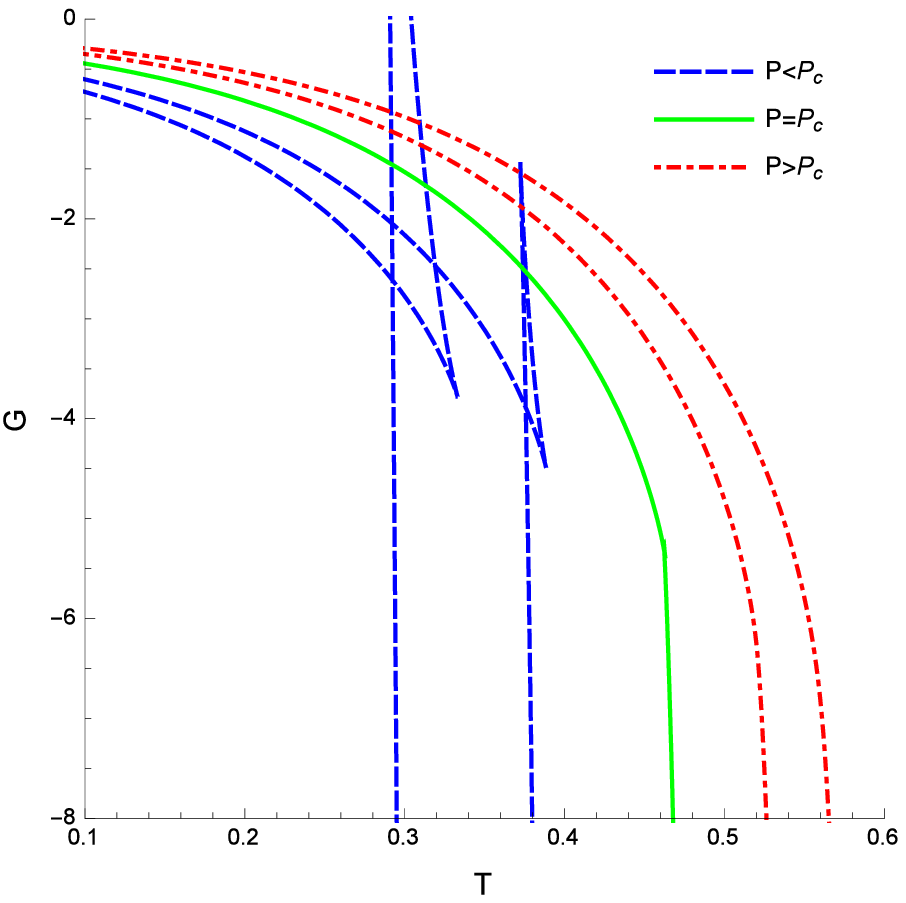}\label{Fig32b}}
	\caption{Critical points: $\kappa=-1$, $\alpha=1$, $\hat{\eta}_{2}=10$, $\hat{\eta}_{3}=-3.5$ and
		$d=10$.
		 There exists a first order phase transition with one physical critical point similar to the Van der Waals system.}
	\label{Figd10}
\end{figure}

\begin{figure}[tbp]
	\centering
	\subfigure[$P-v$ diagram]{\includegraphics[scale=0.6]{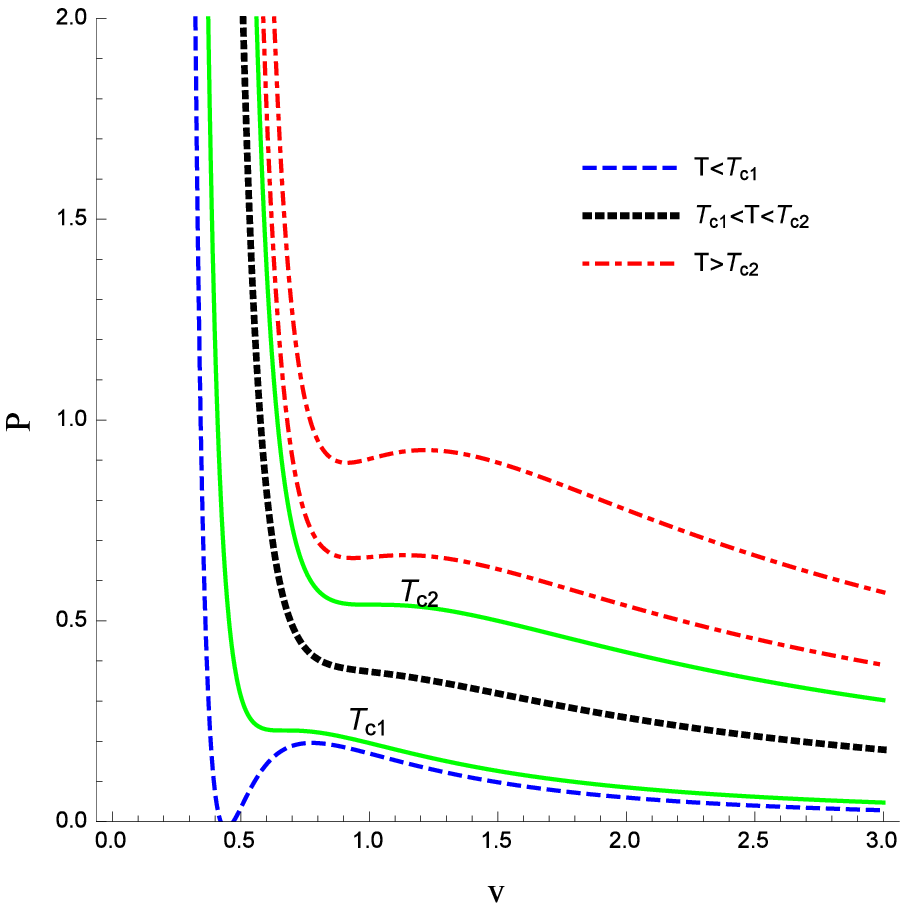}\label{Fig12c}} \hspace*{.2cm}
	\subfigure[Gibbs diagram
	]{\includegraphics[scale=0.6
		]{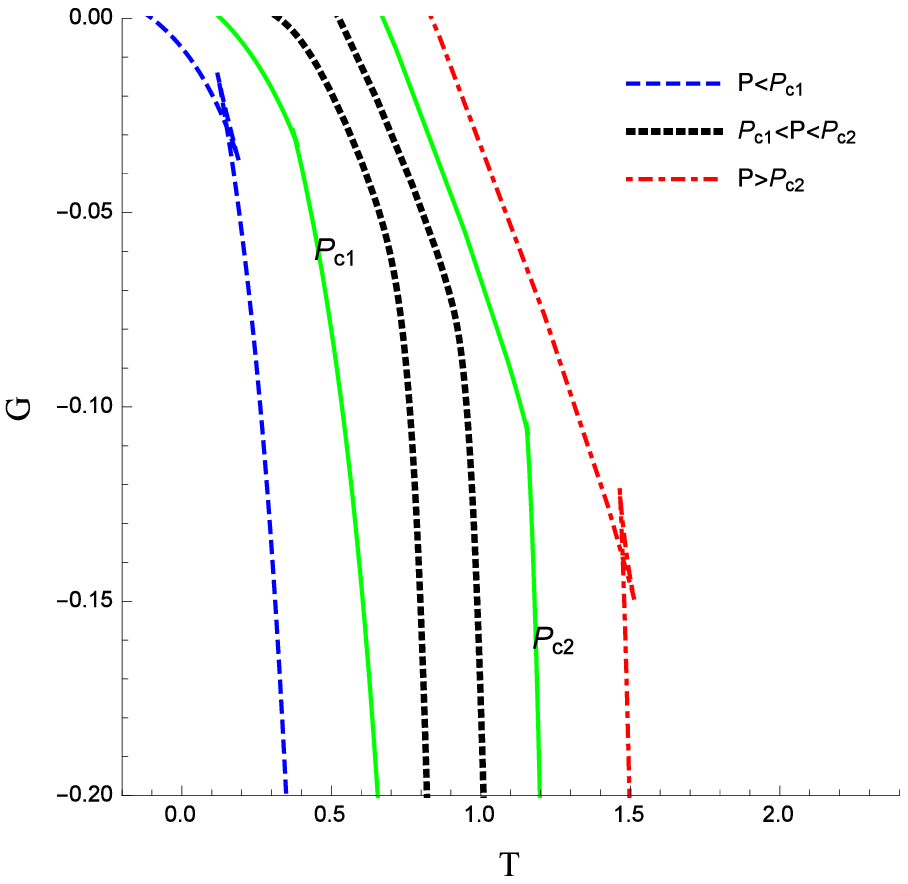}\label{Fig32c}}
	\caption{Critical points: $\kappa=-1$,  $\alpha=1$, $\hat{\eta}_{2}=0.5$, $\eta_{3}=-0.5$ and
		$d=8$.	There are two physical critical point in the above diagrams.
				We have changed the scale of Gibbs diagrams in the way that two critical curves are visible in one diagram. }
	\label{Figd8}
\end{figure}

 In Fig. \ref{Figd10}, we have only one critical point for chosen parameters $d$, $\hat{\eta}_2$, $\hat{\eta}_3$ and $\alpha$. This case is exactly similar to the Van der Waals phase transition with the same isotherm curves. The swallowtail shapes of Gibbs diagras verify the first order phase transition too.
 In Fig. \ref{Figd8}, we choose the parameters so that we can observe two physical critical points. The relevant Gibbs energy in Fig. \ref{Fig32c} shows two critical curves associated with $P_{c1}$ and $P_{c2}$. There is no phase transition for $P_{c1}<P<P_{c2}$. Also the Gibbs energy and its derivatives with respect to the temperature are continuous. We can see first order phase transition for $P<P_{c1}$ or $P>P_{c2}$.

\begin{figure}
	\centering
	\subfigure[ $P-v$ diagram for two of the three critical points]{\includegraphics[scale=0.6]{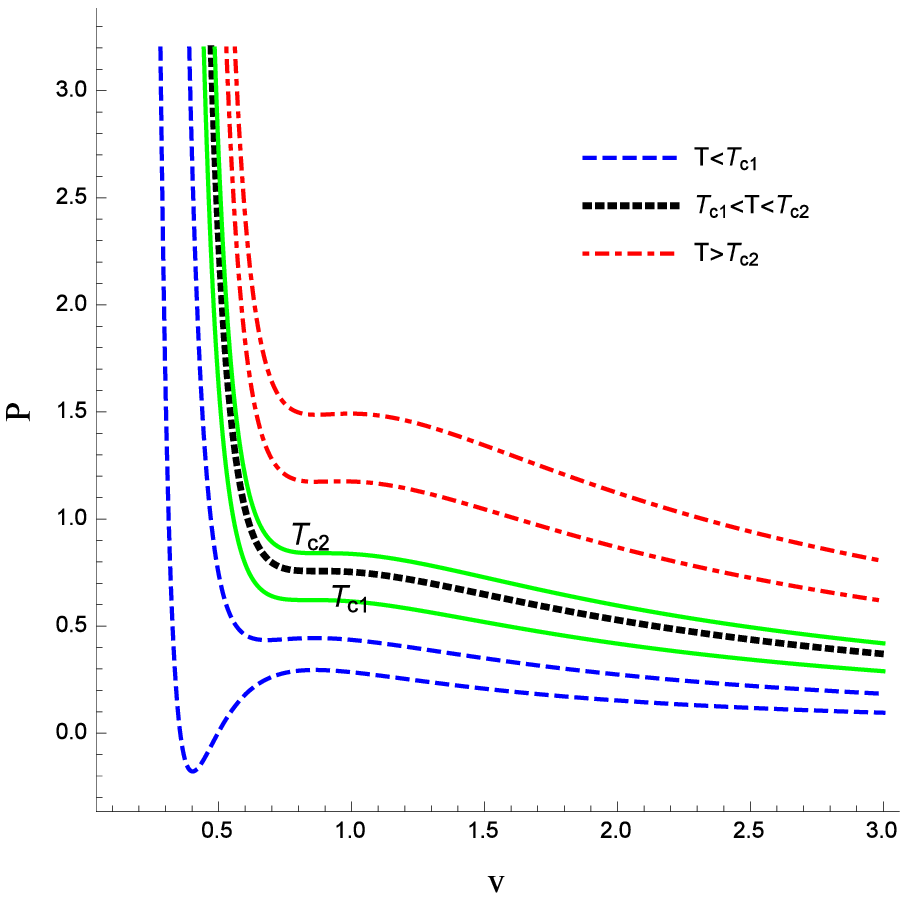}\label{Fig22a}} \hspace*{.4cm}
	\subfigure[Gibbs diagram for two of the three critical points.
	]{\includegraphics[scale=0.6
		]{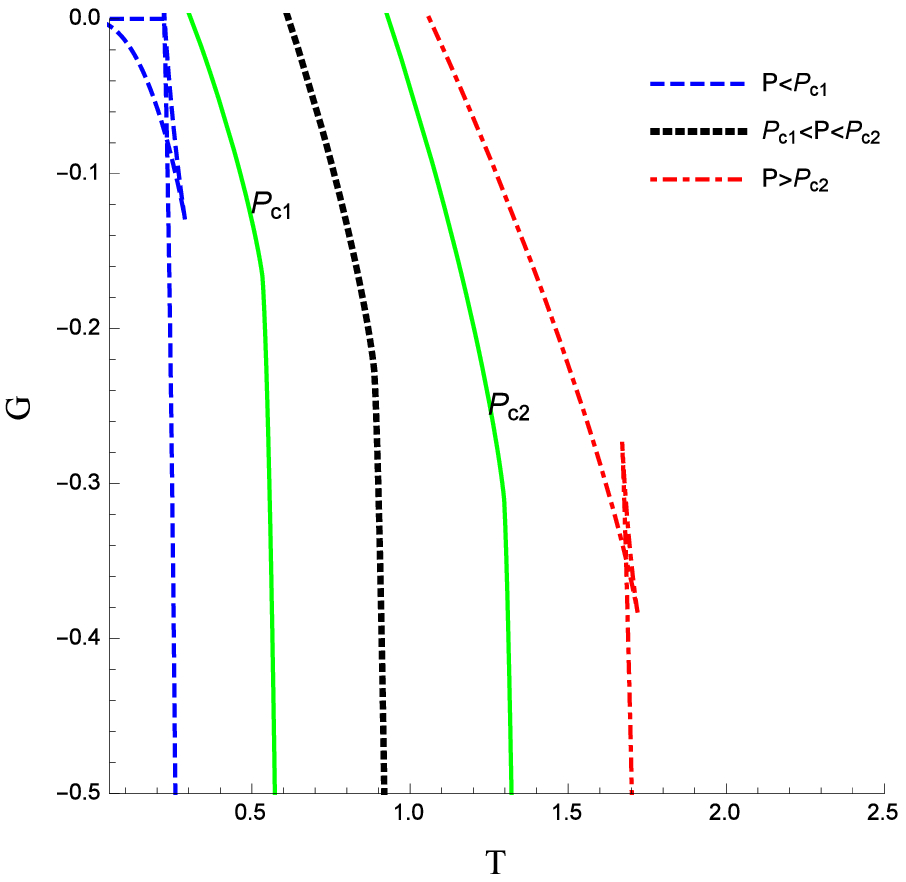}\label{Fig42a}} \hspace*{1cm}
	\subfigure[ $P-v$ diagram for the third critical point.]{\includegraphics[scale=0.6]{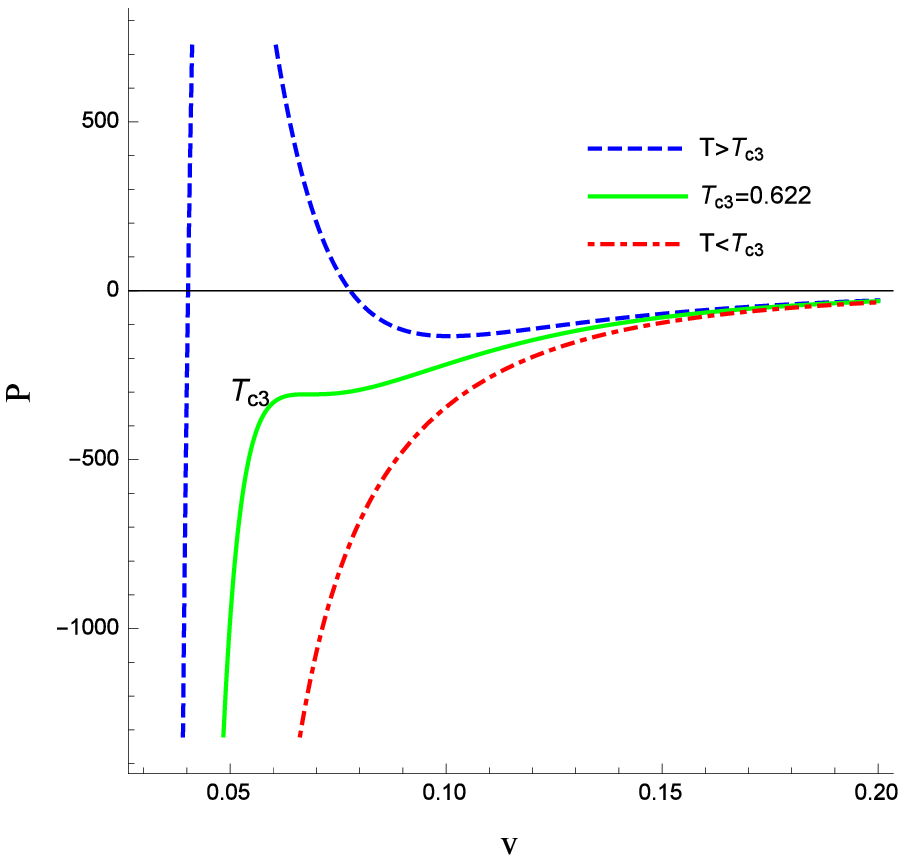}\label{Fig22b}} \hspace*{.4cm}
	\subfigure[Gibbs diagram for the third critical point.
	]{\includegraphics[scale=0.6
		]{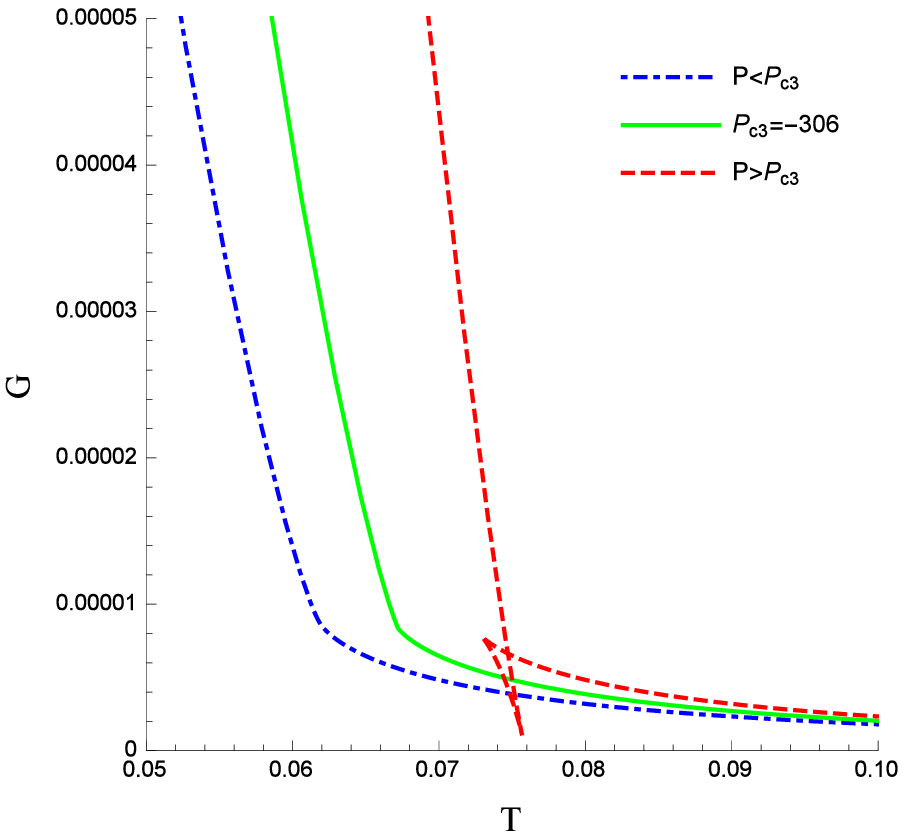}\label{Fig42b}}
	\caption{Critical points:  $\kappa=-1$, $\alpha=1$, $\hat{\eta}_{2}=0.7$, $\hat{\eta}_{3}=3.2$ and
		$d=9$}
	\label{Figd9}
\end{figure}

 The interesting case is the case with three critical points for Lovelock black holes with $\kappa=-1$ which occurs for some specific values of parameters. The corresponding $P-v$ plot is depicted in Fig. \ref{Figd9}.
 Two critical points with positive pressure are shown in Fig. \ref{Fig22a} and the corresponding Gibbs free energy is depicted in Fig. \ref{Fig42a}.
 The third critical point is far from the others and so we bring it in a separate diagram. As it is seen in Fig. \ref{Fig22b}, the third critical point has negative pressure.  Fig. \ref{Fig42b} represents the Gibbs free energy corresponding to this point. It is worth to note that for $P \leq P_{c3}$ the Gibbs energy is completely unphysical with negative compressibility while for $P>P_{c3}$ there exists some physical part in Gibbs diagram. We should emphasis that the curves in Gibbs diagrams are rescaled and shifted for more clarity.

 Following the approach in Sec. \ref{exponent0} and using reduced thermodynamic variables and Taylor expansion for the equation of state, Eq.(\ref{Ph}), we obtain critical exponents as
 \begin{equation}
 \alpha ^{\prime }=0,\quad \beta ^{\prime }=\frac{1}{2},\quad \gamma ^{\prime }=1, \quad \delta ^{\prime }=3
 \end{equation}
 which is consistent with Van der Waals exponents and the results of mean field theory.
 It is worthwhile to emphasise that in a special case with $\hat{\eta}_{2}=\hat{\eta}_{3}=0$, a peculiar isolated critical point emerges for
 hyperbolic black holes and is characterized by non-standard critical exponents, which is discussed in details in Ref. \cite{Frassino}.


\bigskip

   \section{Concluding Remarks}
   \label{conclusion}
    In this study we presented some thermodynamic behaviors of black holes of a more
    general class of solutions possessing non-constant curvature horizons.
    The horizon space of these kinds of black holes is nonmaximally symmetric
    Einstein space. Nontrivial Weyl tensor of such exotic horizons is exposed to the bulk dynamics
    through the higher-order Lovelock term. Investigating the $P-v$ criticality behavior of such
    black holes of Lovelock gravity in the extended phase space
    led to interesting and qualitatively new
    behaviors. By introducing the conjugate quantity to Lovelock parameter $\alpha$,
    We showed that the first law of thermodynamics and the Smarr formula hold.
    By considering the thermodynamics of these kinds of black holes with nonmaximally
    symmetric horizons in cubic Lovelock gravity, we have found some particularly novel
    and interesting results.
    As it is well-known no criticality
    has been found for all known types of Ricci flat black holes in Einstein or Lovelock theories of gravity. Thus
    we went through Ricci flat black holes with nonconstant curvature horizon and we found that
    there exists criticality in every dimension $d>8$ for such black holes with $\kappa=0$. We obtained the
    exact solutions by solving the cubic equation and showed that
    relating to the values
    of chargelike parameters appearing in the metric function, Van der Waals-like behavior and first
    order phase transition may happen. For some values of $\hat{\eta}_{3}$, which is a chargelike parameter
    that is inserted in the metric function due to the appearance of third-order curvature terms,
    two critical point emerge. We have also computed the
    critical exponents of the phase transition and found that in the canonical
    ensemble the thermodynamic exponents coincide with those of the Van der Waals fluid.

    For the black holes with spherical and constant curvature horizons
    critical points do not exist for $d>11$. For our solutions with non constant curvature horizon
    we carried out
    the study numerically and found that one or two critical points exist
    in every dimension even dimensions higher than 11 with the proper choices of the parameters.
    We saw how the value of the parameters that are being emerged in the solutions as a result of the
    nonconstancy of the curvature of the horizon, affect the types of phase transition.
    To disclose the phase structure of the solutions and classify their types, we studied the Gibbs free energy.
    For $\kappa=1$ two different behaviors have been found. For some values of the free parameters,
    a first order phase transition occurs between small and large black holes which is accompanied by
    a discontinuity in the slop of Gibbs free energy at transition point.
    We showed that if the parameters adopt some proper values, a large-small-large black hole
    transition would happen. This process was shown to be accompanied by a finite jump of the Gibbs free energy referred
    to as the zeroth-order phase transition.
    While for the black holes with hyperbolic horizon in different theories
    of general relativity, phase transition is rarely seen, we showed that for our solution in the case $\kappa=-1$,
    in every dimension $d\geq8$, various kinds of interesting phase transitions happen.
    It is interesting enough to see that
    the usual Van der Waals-like small/large black hole phase transition and reentrant phase transition can occur for our solution.
    Also a novel behavior of hyperbolic Lovelock black holes with nonconstant curvature horizon
    has been found for which three critical points could exist with a proper choices of the parameters in the solution.
    Finally it is important to mention that as we carried out the study numerically,
    it is possible to find some other different behaviors by
    other choices of the free parameters.


\end{document}